\documentclass[fleqn,usenatbib]{mnras}

\usepackage{newtxtext,newtxmath}

\usepackage[T1]{fontenc}
\usepackage{ae,aecompl}

\usepackage{graphicx}	
\usepackage{amsmath}	
\usepackage{amssymb}	
\usepackage{xcolor}

\newcommand{\Fermi}{\textit{Fermi}}

\title[GCE and 511 keV from binaries]{Galactic Binaries Can Explain the Fermi Galactic Center Excess and 511 keV Emission}

\author[R. Bartels et al.]{
R. Bartels,$^{1}$\thanks{E-mail: r.t.bartels@uva.nl}
F. Calore$^{2}$, 
E. Storm$^{1}$, 
and C. Weniger$^{1}$
\\
$^{1}$ GRAPPA, Institute for Theoretical Physics and Delta Institute for Theoretical Physics,\\
University of Amsterdam, Science Park 904, 1098XH Amsterdam, The Netherlands \\
$^{2}$LAPTh, CNRS, 9 Chemin de Bellevue, 74941 Annecy-le-Vieux, France
}

\date{Accepted XXX. Received YYY; in original form ZZZ}

\pubyear{2018}

\begin{document}
\label{firstpage}
\pagerange{\pageref{firstpage}--\pageref{lastpage}}
\maketitle

\begin{abstract}
	The \Fermi-LAT Galactic Center excess and the 511 keV
    positron-annihilation signal from the inner Galaxy
    bare a striking morphological similarity. We propose
    that both can be explained through a 
    scenario in which millisecond pulsars
    produce the Galactic Center excess and their
    progenitors, low-mass X-ray binaries, the 511 keV
    signal. As a proof-of-principle we study a specific
    population synthesis scenario from the literature 
    involving so-called ultracompact X-ray binaries.
    Moreover, for the first time, we quantitatively show
    that neutron star, rather than black hole, 
    low-mass X-ray binaries can be
    responsible for the majority of the positrons. 
    In this particular scenario binary millisecond pulsars can
    be both the source of the \Fermi-LAT $\gamma$-ray excess and the
    bulge positrons.
    Future avenues to test this scenario 
    are discussed.
\end{abstract}

\begin{keywords}
binaries -- stars:pulsars -- stars:jets -- gamma-rays:general -- Galaxy:bulge
\end{keywords}




\section{Introduction}
\label{sec:intro}
For the past 8 years, the GeV $\gamma$--ray community has been grasped
by an anomalous signal from the inner-Galaxy present in the \Fermi--LAT data at GeV energies.
First reported by \cite{Goodenough:2009gk}, this so-called Galactic-Centre
excess (GCE) has since been analysed by 
many groups
\citep{Vitale:2009hr, Hooper:2010mq, Hooper:2011ti, Abazajian:2012pn,
Gordon:2013vta, Macias:2013vya, Zhou:2014lva, Daylan:2014rsa, Calore:2014xka, 
TheFermi-LAT:2015kwa, Bartels:2017vsx}.
The GCE generated a lot of excitement,
because
the signal spectrum and morphology are consistent with that expected
from a typical weakly-interacting massive particle dark matter (DM)
candidate \citep[e.g.][]{Hooper:2010mq, Calore:2014nla}.
Also, its morphology appears consistent with that of annihilating DM, 
being well
fit by a spherically-symmetric NFW \citep{Navarro:1996gj} profile with 
an inner slope of $\gamma=1.26$.
However, alternative astrophysical scenarios have since been proposed, 
the most promising of
which is a population of unresolved millisecond pulsars (MSPs)
\citep{Abazajian:2010zy, Gordon:2013vta, Yuan:2014rca, Abazajian:2014fta, 
Petrovic:2014xra}. \cite{Bartels:2015aea} and \cite{Lee:2015fea} found corroborative evidence for such an unresolved population of point sources. 
If the GCE is generated by MSPs
they would have a roughly radially-symmetric $r^{-2.5}$ density profile.
Finally, in a recent paper we reanalyzed the GCE using the newly developed
$\gamma$--ray fitting code \texttt{SkyFACT} \citep{Storm:2017arh} and
found that a boxy-bulge model describes 
the morphology well, providing further evidence for a stellar origin of the GCE \citep{Bartels:2017vsx}. 
Similar results have been obtained by \cite{Macias:2016nev}.
In the near future, this scenario can be tested by the MeerKAT and
SKA radio telescopes \citep{Calore:2015bsx}.

Another open question in astroparticle physics is the origin of the positrons
producing the 511 keV line
emission from the Galactic bulge. 
The signal has been around for 
four decades \citep{1978ApJ...225L..11L}, and has been accurately studied in the past
15 years with the Spectrometer aboard INTEGRAL (SPI) 
\citep[e.g.~][]{Knodlseder:2005yq,Siegert:2015knp}. What is surprising about the 511 keV
emission from the 
Milky Way is the high bulge-to-disk ratio.
Early analyses suggested the 
flux ratio to be $B/D\sim 1\text{--}3$ \citep{Knodlseder:2005yq}, however, 
more recent analyses find smaller values $B/D<1$~\citep{Bouchet:2010dj, Siegert:2015knp}.
Many sources for the positrons have been proposed.
First, radioactive isotopes produced during stellar nucleosynthesis inject positrons into the interstellar medium (ISM) \citep[e.g.~][]{1973NPhS..244R.137C}.
There is a guaranteed contribution from 
$^{26}$Al and $^{44}$Ti produced by nucleosynthesis in massive
stars and core-collapse supernovae. The synthesis of these elements yields about $\sim 3\times 10^{42}\mathrm{\,e^+\,s^{-1}}$ each. Depending on the assumed B/D luminosity ratio this contributes $\mathcal{O}(10\%-100\%)$
to the disk 511 keV flux, where $\sim 10\%$ is closest to the most recent estimates.
However, these guaranteed $^{26}$Al and $^{44}$Ti sources are insufficient to explain the 
emission from the bulge \citep{Alexis:2014rba,Siegert:2015knp}.
Second, astrophysical sources such as for instance
the super-massive black hole in the center of our Galaxy
Sagittarius A* (Sgr A*) \citep{Totani:2006zx}, pulsars 
\citep{Wang:2005cqa} or 
Galactic X-ray binaries (XRBs) can also 
produce positrons
\citep{Guessoum:2006fs, Bandyopadhyay:2008ts}. 
Finally, the positrons could come from
exotic sources such as
annihilating MeV DM \citep{Boehm:2003bt} or 
de-exciting DM \citep{Finkbeiner:2007kk}
\footnote{The MeV--DM model is by
now ruled out by cosmological constraints 
\citep{Wilkinson:2016gsy} and also de-exciting DM starts to be constrained
\citep{Frey:2013wh}.}.
For a complete review of the 
511 keV signal see \cite{Prantzos:2010wi}.

The GCE and the bulge component of the $511\mathrm{\,keV}$ emission show a few
strikingly common features: they both are roughly spherically symmetric,
extend $\sim 10^\circ$ away from the Galactic center (GC), 
are consistent with a uniform spectrum, and strongly peak towards the GC. 
Within the available analysis uncertainties, 
these similarities are suggestive of a common origin.
This possibility has been explored in recent years by a few theoretical works.
\cite{Boehm:2014bwa} considered a common origin in terms of dark matter.
More recently, \cite{Crocker:2016zzt} proposed $^{44}$Ti produced in thermonuclear
supernovae as the source of positrons in the bulge. Their stellar-evolution model
also predicts a population of millisecond pulsars.

In this paper we propose a new connection of the
511 keV emission and GCE through
population synthesis of low-mass X-ray binaries
(LMXBs).
LMXBs are a particular type of X-ray binaries
consisting of a compact object accretor 
and a low-mass companion star ($\lesssim 1\mathrm{\,M_\odot}$).
The
accretor is either a neutron star (NS) or
a black hole (BH).
LMXBs with NS accretors are
MSP progenitors, via the standard "recycling" scenario
\citep{1991PhR...203....1B, 2007A&A...469..807L}.
Consequently, if MSPs are responsible for the GCE, LMXBs are expected to be present in considerable
numbers in the Galactic bulge. 
In fact, LMXBs were already suggested as a potential candidate to
explain the 511 keV emission because of their
distribution being concentrated towards the 
bulge~\citep{2004ESASP.552...15P}. 
However, 
this scenario was excluded because the brightest LMXBs
(in X-rays) lie mostly in the disk and thus cannot reproduce the
correct morphology if the positron-injection rate scales with X-ray flux~\citep{2004ESASP.552...15P}.
Nevertheless, X-ray binaries can power jets 
causing radio emission which is correlated with the X-ray brightness
\citep[e.g.~][]{Remillard:2006fc}.
Binaries with compact object accretors were
proposed as the source of low-energy positrons by~\cite{Guessoum:2006fs} and \cite{Bandyopadhyay:2008ts}. 
These works, however, focused mostly on BH accretors. 
The microquasar interpretation from \cite{Guessoum:2006fs}
has recently gained support from the claimed detection
of annihilating positrons from 
V404 Cygni by \cite{Siegert:2016ymf}
\citep[but also see][]{Roques:2015bma}.
We propose a model capable of explaining both the GCE and 511 keV line
through the evolution of a subclass of LMXBs
called ultracompact X-ray binaries (UCXBs). UCXBs could generate  
a population of MSPs responsible for the $\gamma$-ray GCE emission and
sustain the positron-production rate required for the 511 keV line through jet emission from NS- and/or BH-LMXBs.
For the first time, we show that NS-LMXBs alone can 
be responsible for the majority of the positrons.
These objects were typically discarded as the main source of the 511 keV line 
emission because, 
as a result of the traditionally assumed scaling relation between radio and X-ray emission and its extrapolation down to the quiescent regime,
their jet emission was expected to be subdominant to that of BH-LMXBs.
We instead show that, under reasonable assumptions on the jet physics,
jet emission from NS-LMXBs can in fact be important in the quiescent state.

The paper is organized as follows: 
in Sect.~\ref{sec:bin_pos} we discuss X-ray binary jets and their potential positron yield.
Next, we discuss a population synthesis model based on 
UCXBs in the bulge in Sect.~\ref{sec:pop_synth}.
In Sect.~\ref{sec:prospects} we discuss various future lines of research that can be
pursued to test the connection between the 511 keV line and the GCE and to study
whether LMXBs can be responsible for the majority of positrons in the Galactic
bulge.
We discuss our results and conclude in Sect.~\ref{sec:discussion}.
%
%
\section{Positrons from Galactic binaries}
\label{sec:bin_pos}
Jet emission from Galactic binaries is a potential source of positrons
in our Galaxy. \cite{Guessoum:2006fs} and \cite{Bandyopadhyay:2008ts} 
proposed BH-LMXBs as the source of the Galactic
bulge positrons. NS-LMXBs also seem capable 
of launching jets similar to BHs \citep{Mirabel:1999fy}. 
Despite their larger number compared to BHs,
NS-LMXBs have typically been given less
attention in context
of the positron excess in the bulge, primarily
because their jets in the quiescent state are expected to be weaker \citep{Fender:2005nh, Bandyopadhyay:2008ts}.
Below, we will discuss the positron yield from Galactic binaries, both with NS- and BH-accretors.

\subsection{Jets from black holes}
\subsubsection{Accretion}

Accreting BHs exhibit different luminosity and spectral states that depend on the observed X-ray emission. In the so-called high/soft state, the X-ray emission is bright with a soft spectrum, and is thought to originate from thermal emission from the accretion disk itself. Here, the accretion rate onto the BH is also high. In the low/hard state, the X-ray emission is fainter but the spectrum is hard; the X-ray emission is likely nonthermal, possibly originating from the corona, although the origin is not clear \citep{Remillard:2006fc, Done:2007nc}. The accretion rate is low in this state. In the low/hard state, accreting BHs can launch jets, and this state is also sometimes referred to as the jet-dominated state. Radio emission is also observed in the low/hard or jet-dominated state, which is likely due to synchrotron radiation from the jet. The quiescent state is often interpreted as the low/hard state at even lower accretion rates, resulting in even dimmer X-ray emission.

BH-LMXBs typically have quite low accretion rates, $\mathcal{O}(1\%\,\dot{M}_\mathrm{Edd})$. At these accretion rates a jet
is expected to be present \citep{Fender:2003ae}.
For BH-XRBs binaries
in the low/hard state there exists a tight empirical relation between the X-ray emission arising from the disk and the radio emission from the jet:
\begin{equation}
  \label{eq:radioX}
  L_R \propto L_X^\beta\;,
\end{equation}
with $\beta \sim 0.7$
\citep{Corbel:2003wt, Gallo:2003tv, Gallo:2014lba}.
This relation has been observed down to X-ray luminosities of 
$L_X = 10^{-8.5}\,L_\mathrm{Edd}$ \citep{Gallo:2006dz, Gallo:2014lba}, 
far into the quiescent regime.\footnote{
In the literature, quiescent typically means 
$L_X\lesssim 10^{-5}\,L_\mathrm{Edd}$.}
The radio/X-ray correlation in BH-XRBs has later been extended
into a broader, empirically founded, 
unified accretion model of
inefficiently accreting BHs across all masses, from AGNs down to Galactic BH-XRBs, known as the fundamental plane \citep{Merloni:2003aq, Falcke:2003ia}.

For BHs in the low/hard state, most of the radiated energy is in the hard X-ray band, $\gtrsim10\mathrm{\,keV}$.
The 0.1--200 keV X-ray flux has been
taken as a good estimate of the bolometric luminosity.
The bolometric luminosity of an accreting BH follows
from the conversion of gravitational potential energy into
radiation (for more details on the connection between the bolometric luminosity and
the accretion rate see Appendix \ref{sec:luminosity}):
\begin{align}
\label{eq:Lbol}
  \begin{split}
	L_\mathrm{Bol} &= \eta \dot M c^2\;.
  \end{split}   
\end{align}
Here, $\eta$ is the radiative efficiency, which is given
by:
\begin{equation}
\label{eq:eff}
\eta = \left\{
\begin{split}
  &0.1 \qquad &\text{for }\dot M > \dot M_\mathrm{crit} \\
  &0.1 \dot M / \dot M_\mathrm{crit} \qquad
  &\text{for }\dot M\leq\dot M_\mathrm{crit}
\end{split}
\right.\;,
\end{equation}
with $\dot M_\mathrm{crit} = 0.01\dot M_\mathrm{Edd}$
for BHs \citep{Fender:2003ae}.
Systems in the low/hard state have
$\dot M\leq\dot M_\mathrm{crit}$ \citep{Narayan:1994xi, Kording:2006sa}. 
The broadband spectrum follows $S_\nu \propto \nu^\alpha$
with $\alpha = -0.6\,(0)$ for X-ray (radio) emission.%
\footnote{For an
example of a spectrum of a BH in the low/hard state see for example Fig.~1 in \cite{Markoff:2000ir}. Also see \cite{Done:2007nc} for a discussion of the different spectral states.}
In this case, we find $L_X[2\text{--}10\mathrm{\,keV}] \sim 0.15 L_\mathrm{Bol}$\footnote{
The conversion is most sensitive to the high energy cutoff of the X-ray emission due to its hard spectrum.}.

\subsubsection{Black-hole jet power}
Since jets are present in the low/hard state we
will henceforth only refer to inefficiently accreting
BHs.
The radio brightness of such sources scales with jet power as $L_R\propto L_J^{1.4}$ \citep{Blandford:1979za, Falcke:1994eb, Markoff:2002xs, Heinz:2003xt}.
Consequently, 
it follows from the empirical scaling between radio and X-ray luminosity that the jet power scales with the
square root of the X-ray luminosity \citep{Fender:2003ae}:
\begin{equation}
\label{eq:Pj_BH}
  L_J = A_\mathrm{BH} L_{X}^{0.5},
\end{equation}
where $L_{X}$ and $L_J$ are given in Eddington units.
In what follows we will use $L_\mathrm{Bol}$ instead of 
$L_{X}$ since the hard X-ray luminosity is a proxy for the
bolometric luminosity as discussed above.
The prefactor, $A_\mathrm{BH}$, relates the jet kinetic luminosity 
(i.e.~the kinetic energy carried by matter, excluding the energy in
magnetic field)
to the radiative luminosity. We assume that the jet is steady, i.e.~injecting energy
into the ISM at a constant rate. 
The parameter $A_\mathrm{BH}$ can be estimated using
calorimetric arguments by looking at
bubbles traced-out by the shocked ISM \citep{Gallo:2005tf}. 

A conservative lower limit of 
$A_\mathrm{BH}\gtrsim 6\times10^{-3}$
was derived by \cite{Fender:2003ae}.
\cite{Gallo:2005tf} study the jet in Cygnus-X1 and find that at $L_{\mathrm{Bol}} = 0.02 L_\mathrm{Edd}$, $L_J = \left(0.06\text{--}1\right) L_{\mathrm{Bol}}$.
Using Eq.~(\ref{eq:Pj_BH}) this translates into 
$A_\mathrm{BH}=8\times10^{-3}\text{--}0.14$.
These values are also in agreement with the findings of \cite{Heinz:2005hw}. 
Altogether, an appropriate range is $A_\mathrm{BH}\in\left[6\times10^{-3},0.3\right]$ 
\citep{Fender:2005nh}. 
We will  use a benchmark value of
$A_\mathrm{BH}=0.1$.
Note that this value of $A_\mathrm{BH}$ also corresponds
to a transition from the high/soft to the low/hard
state at $0.01\mathrm{\,\dot M_\mathrm{Edd}}$ following
\cite{Fender:2003ae}.

The jet power can also be expressed
in terms of the mass accretion rate. 
For an inefficiently accreting
BH it becomes
\begin{align}
\label{eq:Pj_Mdot}
  \begin{split}
	L_J &= A_\mathrm{BH} \left(\eta \dot M c^2\right)^{0.5} \\
  		&\approx 5.67\times10^{35}
        \left(\frac{A_\mathrm{BH}}{0.1}\right)
        \left(\frac{\dot M}
        {10^{-10}\mathrm{\,M_\odot\,yr^{-1}}}\right)
        \mathrm{\,erg\,s^{-1}},
  \end{split}
\end{align}
where we switched from Eddington to cgs units in the second line.

\subsection{Jets from neutron stars}
\label{sec:NS}
Similar to BH X-ray binaries, NS X-ray binaries also 
have hard states at low accretion rates
\citep{Migliari:2005hv}. During
these states they can also have steady jets.
However, for the same X-ray luminosity NSs
are less radio loud and thus their jets are 
supposedly weaker. In addition, 
the 
radio--X-ray scaling in Eq.~(\ref{eq:radioX}) for 
many NSs is 
different from that in
BHs, with $\beta=1.4$ in the hard-state \citep{Migliari:2002tc}.
This implies
$L_J \propto L_X$; if this correlation holds down
to low mass-accretion rates then NSs
would have negligible jet power compared to BHs and
never be in a jet-dominated state \citep{Migliari:2002tc}.
However, there are hints that at the lowest mass-accretion
rates NS-LMXBs do become jet dominated and that the radio X-ray correlation has a slope $\beta\sim 0.7$, similar to what is observed in BHs \citep{Deller:2014yca}. In this case one also
gets a relation:
\begin{equation}
\label{eq:Pj_NS}
  L_J = A_\mathrm{NS} L_{X}^{0.5}.
\end{equation}
One class of NS binaries in particular display clear indications for them launching jets in the low/hard and quiescent regimes: the transitional millisecond pulsars (tMSPs) \citep{Deller:2014yca}. These are sources that transition between a radio pulsar state and an LMXB state \citep{Archibald:2009zb}.
This class of sources is also in a low/hard-state and two
of the four observed sources are particularly dim in X-rays, dimmer than any other NS-LMXB used to constrain the radio--X-ray correlation \citep[see Fig.~10 in][]{Deller:2014yca}. The three
confirmed tMSPs have been observed to lie on a $\beta\sim 0.7$ track in the radio--X-ray correlation \citep{Deller:2014yca}. In the LMXB state there are hints of jet emission, for instance in J1023+0038, the dimmest of the tMSPs. The collimated outflow is assumed to arise from the propeller mechanism~\citep[e.g.,][]{Romanova:2005di} in which a centrifugal barrier is formed preventing accreted material from reaching the NS surface and instead being expelled outwards. This propeller, and the resulting jet, is expected to be on for about 20\% of the time when in the LMXB state \citep{Archibald:2014nda, Jaodand:2016hry}.
The radio luminosity is only a factor $\sim 5$ less bright than that of BHs at the same X-ray luminosity. Using the radio loudness as a proxy for the jet power the ratio of NS-to-BH jet power is (Eq.~\ref{eq:radioX})
$\frac{L_{J,\,\mathrm{NS}}}{L_{J,\,\mathrm{BH}}} = \left(\frac 1 5 \right)^{1/1.4}\approx 0.3$, resulting in a scaling $A_\mathrm{NS} = 0.3 A_\mathrm{BH}$ \citep{Fender:2003ae,Deller:2014yca}. 
In terms of the mass accretion rate the jet power of the NSs assuming a $\beta=0.7$ scaling
is: 
\begin{align}
\label{eq:Pj_Mdot_NS}
  \begin{split}
	L_J &\approx 5.38\times10^{35}
        \left(\frac{A_\mathrm{NS}}{0.03}\right)
        \left(\frac{\dot M}
        {10^{-10}\mathrm{\,M_\odot\,yr^{-1}}}\right)
        \mathrm{\,erg\,s^{-1}}.
  \end{split}
\end{align}
The similarity between the prefactor in Eq.~(\ref{eq:Pj_Mdot_NS}) and the  
one in Eq.~(\ref{eq:Pj_Mdot}) is not a coincidence, 
but it is because we use
$\dot M_\mathrm{crit}\approx A^2 \dot M_\mathrm{Edd}$
and because $\eta$ is the same for NSs and
BHs.\footnote{
Using 	$\dot M_\mathrm{crit}\equiv A^2 \dot M_\mathrm{Edd}$ Eqs.~(\ref{eq:Pj_Mdot}) and (\ref{eq:Pj_Mdot_NS}) become
$L_J = 5.67\times10^{35}
\left(\frac{\dot M}
{10^{-10}\mathrm{\,M_\odot\,yr^{-1}}}\right)
\mathrm{\,erg\,s^{-1}}$.}
In this scenario, jets in NSs and BHs are identical for the same mass-accretion rate. We can rewrite Eqs~(\ref{eq:Pj_NS}--\ref{eq:Pj_Mdot_NS}) and (\ref{eq:Pj_BH}--\ref{eq:Pj_Mdot}) as $L_J = 0.1 \dot M c^2$ in {\bf cgs} units, where the 0.1 comes from $\eta$.

In this work, we use the radio--X-ray correlation from tMSPs
as our reference case for
NS-LMXB jets. We use a population-synthesis model of UCXBs and
assume that these systems have similar jet properties
to tMSPs. Although there are currently no UCXBs that have been identified as tMSPs, the UCXBs in our sample
have very small mass-accretion rates, comparable to that observed 
in tMSPs. This justifies the assumption of similar jets.
We will assume a duty cycle of the jet of 20\%. The assumed duty cycle is based on the fraction of the time the propeller mechanism is on, but this is somewhat optimistic since it is unclear what fraction of the time the source spends in the LMXB state where an accretion disk is present.
We take $A_\mathrm{NS}=0.3 A_\mathrm{BH}$.
Following \cite{Fender:2003ae}, this value of $A_\mathrm{NS}$ means that the transition from the high/soft state
to the low/hard state occurs at an Eddington rate which is about a
factor 10 below the rate at which this transition occurs for BHs, 
i.e.~$\dot M_\mathrm{crit} = 10^{-3} \dot M_\mathrm{Edd}$.
We will assume no jets are present when 
$\dot M >\dot M_\mathrm{crit} = 10^{-3}$.
Finally, we will use a conversion
$L_X[2\text{--}10\mathrm{\,keV}] \sim 0.4 L_\mathrm{Bol}$,
which is reasonable given the spectral index of $1.17$ 
and the uncertainty
in the cutoff of the X-ray spectrum
of J1023+0038 \citep{Tendulkar:2014wga, Archibald:2014nda}.
We note that $L_X[2\text{--}10\mathrm{\,keV}]$ is only used for 
plotting purposes and serves as a reference of the X-ray flux.
Our estimate for the positron yield does not depend on this conversion.

\subsection{Positrons from jets}
To calculate the positron yield from Galactic binaries
we need to know the number of positrons injected into the ISM by the jet. 

The composition of jets remains largely unconstrained.
A first open question is whether the kinetic energy in the jet is carried away 
by hadrons or by an electron-positron plasma
\citep[e.g.~][]{Migliari:2002tc, Chattopadhyay:2004gf, Romero:2005fr}.
Other uncertainties 
are related to the Lorentz factors of the jet bulk matter and to that
of the plasma. However, the observed radio synchrotron emission means there must be
high-energy electrons and magnetic fields present
\citep[e.g.~][]{Heinz:2002qj}. 

In what follows we will assume that the jets are composed of an 
electron-positron plasma, which makes our estimates 
optimistic. Moreover, we will assume that steady jets are only mildly
relativistic, with a bulk Lorentz factor 
$\Gamma_\mathrm{jet}\leq 1.4$ \citep{Fender:2004gg}.
For transients the Lorentz factor can be much higher,
$\Gamma_\mathrm{jet}\geq 2$ \citep{Fender:2004gg,Remillard:2006fc},
however, we do not consider transient jets in this work. 
Finally, we assume that the jet plasma is cold -- that is, the majority 
of the energy is carried by non-relativistic particles with a mean
Lorentz factor $\left<\gamma\right>\sim 1$ in the jet rest frame. 
Although these low-energy particles
are unobservable -- since they do not produce synchrotron emission --
a small bulk velocity and a predominance of cold particles in the plasma
are required if binaries are 
primarily responsible for the injection of positrons into the bulge. 
\cite{Beacom:2005qv} and \cite{Sizun:2006uh} showed that if 
the positrons responsible for the 511 keV line emission
are injected into the ISM with energies much above $\gtrsim 3\mathrm{\,MeV}$
they should also produce observable continuum emission due the 
direct annihilation in-flight with electrons at rest, which produces a boosted spectrum instead
of the narrow line from annihilation through 
positronium formation.
Consequently, both the $\Gamma_\mathrm{jet}$ and 
$\left<\gamma\right>$ have to be
sufficiently small to ensure that the injected leptons have small momenta
in the rest frame of the ISM. We note that annihilation
inside the jet, rather than the ISM, would lead to a large Doppler shift and broadening of the line. Spectroscopy of the line 
constrains bulk motions of the medium in which the positrons
annihilate to be $\sim 100\mathrm{\,km\,s^{-1}}$ with respect
to the Earth frame \citep{Siegert:2015knp}.
Although hypothetical, this scenario is not unreasonable given our understanding of jet physics \citep{Heinz:2002qj, Fender:2004gg}. 
Following \cite{Heinz:2002qj} the injection rate of 
positrons under the above assumptions is given by:
\begin{align}
\label{eq:Npos}
  \begin{split}
  \dot N_{e^+} &= \frac{L_J}
  {2 \left<\gamma\right> \Gamma_\mathrm{jet} m_e c^2} \\
  &\approx 4.36\times 10^{40}
  \left(\frac{\left<\gamma\right>}{1}\right)^{-1}
  \left(\frac{\Gamma_\mathrm{jet}}{1.4}\right)^{-1}
  \left(\frac{L_J}{10^{35}\mathrm{\,erg\,s^{-1}}}\right)
  \mathrm{\,s^{-1}}.
  \end{split}
\end{align}
We caution that for certain BHs this exact modeling cannot be correct. For instance, the high-mass X-ray binary Cygnus X-1 should be detectable
as a 511 keV point source with INTEGRAL/SPI if modeled using the above assumptions. Also the supermassive BH at the center of our Galaxy, Sagittarius A* (Sgr A*), would have a positron yield larger than all Galactic binaries combined. We comment
on these sources in Appendix \ref{sec:individual_sources}.

\subsubsection{511 keV emission}
Assuming that all positrons annihilate in the vicinity of the source
and a fraction $f_p$ 
of the low-energy positrons form positronium,
we can calculate the total 511 keV photon production rate:
\begin{align}
\label{eq:N511}
  \begin{split}
  \dot N_{511} &= \frac{2 f_p \dot N_{e^+}}{4}
  \mathrm{\,s^{-1}},
  \end{split}
\end{align}
where the numerical prefactors arise because 75\% of time the ortho-positronium (o-Ps)
bound state is formed
which annihilates to 3 photons and only in 25\% of the cases para-positronium (p-Ps)
is formed which annihilates to two mono-energetic photons 
\citep[e.g.~][]{Prantzos:2010wi}. The positronium fraction $f_p$ is set to 1
in accordance with spectroscopic studies of the line \citep{Siegert:2015knp}.

\medskip


\section{Population synthesis}
\label{sec:pop_synth}
If MSPs are responsible for the GCE, as suggested by recent analyses
\citep{Bartels:2015aea, Lee:2015fea, Caron:2017udl},
they must
have either evolved in-situ or have been injected into the Galactic bulge
\citep[for instance through disrupted globular clusters][]{Brandt:2015ula, Arca-Sedda:2017qcq, Fragione:2017rsp}.
In either case, progenitors of MSPs are expected to be, or have been,
present in the bulge.

In this work, we will assume these progenitors to be X-ray binaries
with a low-mass companion, i.e.~LMXBs, where MSPs arise as recycled NSs, spun-up by accreting
matter and angular momentum from their companion \citep{1991PhR...203....1B}. 
As a side remark, we mention the possibility that, under certain circumstances, no LMXB phase has to be present
before MSP formation, as noted by \cite{Ploeg:2017vai}. 
In the case of accretion-induced collapse of a white dwarf
the NS created can be directly spun-up. This is how, for example, MSPs arise in the
scenario of \cite{Crocker:2016zzt}.

\textit{All roads lead to Rome.} There is a plethora of 
evolutionary paths to form MSPs from LMXBs.
Various formation and evolutionary channels of 
X-ray binaries in globular clusters are studied by \cite{Ivanova:2007bu}.
Figure 1 of that work depicts the scenarios to form MSPs in globular
clusters along with the properties of the system, such
as whether the system is in a binary or isolated and the lifetime of the MSP.
In fact, the authors mention that they end up with an overproduction of MSPs
if all these channels efficiently lead to MSPs. 
As such, from an evolutionary perspective,
the origin of the MSPs producing the GCE is still largely unknown. 

Given that the precise evolutionary track is hard to constrain, the goal of the present work is not to present a complete overview or
in-depth modeling of the 511 keV and GCE symbiosis, but is instead to
propose a viable population synthesis scenario which could explain simultaneously 
the observed positron injection rate and $\gamma$--ray emission from the 
Galactic bulge. 
In this scenario, NS- and BH-LMXBs produce the low-energy
positrons, while the NS-LMXBs are also responsible for the MSP 
population leading to the $\gamma$--ray emission.
Such a scenario is mainly a proof-of-principle and can
serve to highlight new lines of research to test a common origin of the
511 keV line and the GCE, some of which are provided in Sect.~\ref{sec:prospects}.

\subsection{A model: ultracompact X--ray binaries}
\subsubsection{Setup}
UCXBs are a subclass of
LMXBs with short periods, $<1\mathrm{\,h}$ for
the brightest systems, implying 
a small separation between the accretor and the donor star. The 
donor is typically a low-mass star such as a white dwarf or helium star.
Due to gravitational-wave emission from the binary, the orbit shrinks
until the donor overflows its Roche lobe resulting in mass transfer onto 
the companion, which can be a NS or a BH. Mass transfer leads to an overall mass loss of the system and therefore a widening of the orbit
in order to conserve angular momentum. The wider orbit results
into a corresponding increase in the rotation period, less gravitational
wave emission and a decreasing mass transfer rate \citep{vanHaaften:2011iy}.
\cite{vanHaaften:2013uuo,vanHaaften:2015wta} predict that UCXBs
are about 100 times more common in the Galactic bulge today than 
regular LMXBs.
However, due to their low mass-accretion rates today most UCXBs
would not be visible in current X-ray surveys, which typically are
sensitivity limited at $\gtrsim 10^{31}$--$10^{32}\mathrm{\,erg\,s^{-1}}$ in
the $2$--$10$ keV band 
\citep{Jonker:2011gf, Hong:2016qjq, Zhu:2018pkm}.
Their large numbers make UCXBs a
promising candidate to be the dominant channel for MSP formation in
the bulge. Moreover, the UCXB X9 in the Globular cluster Tucanae 47
(47 Tuc X9) potentially is a tMSP \citep{Bahramian:2017mbs}.

\cite{vanHaaften:2013uuo} model the expected population of UCXBs
in the Galactic bulge today. They find that a population 
of up to $\sim 2\times10^{5}$ UCXBs with NS accretors (NS-UCXB)
should be present in the bulge
and stress 
that such a population is necessary to match the number of
detected bright bulge sources. The bright sources also have short orbital periods ($<1\mathrm{\,h}$). Due to the fast evolutionary
timescale most binaries will have longer
periods, $>1\mathrm{\,h}$, and correspondingly are less bright and undetectable, hence the large number of binaries required to match observations
\citep{vanHaaften:2011iy,vanHaaften:2012qc,vanHaaften:2013uuo}.
The NS-UCXBs can form MSPs,
which could be isolated
if they evaporate the donor star or be present
as a binary system \citep[also see][]{vanHaaften:2012zn}. 
The number of MSPs present in the bulge today resulting from 
NS-UCXBs could therefore be $\mathcal{O}(10^5)$ if they
have not spun down yet.
This number is qualitatively in
agreement with the number required to explain the 
GCE\footnote{The total number of MSPs required to explain the GCE
depends critically on the $\gamma$--ray luminosity function
and can be as low as $\mathcal{O}(\mathrm{few}\times100)$ up to much
larger values by orders of magnitude, depending on the minimum luminosity and the shape of the luminosity function.}
\citep{Bartels:2015aea,Fermi-LAT:2017yoi, Bartels:2017xba}
and we will
assume that these MSPs are indeed responsible for the GCE.

No UCXBs with BH accretors form in the simulation of \cite{vanHaaften:2013uuo}, 
since binary systems with massive stars collapsing
directly into a BH tend to merge before they can evolve upto
the point of mass transfer. However, the UCXB 
population-synthesis model
of \cite{Belczynski:2003ac} is consistent
with a ratio BH-to-NS of 1:5. 
In this model the BHs form through
accretion-induced collapse of the NS.
In addition, BH-LMXB systems (or candidates), identifiable through
the ratio of hard X-ray emission to radio emission, 
have been discovered both in extragalactic globular clusters
~\citep[e.g.][]{2007Natur.445..183M,2011MNRAS.415.1875M}
as well as in Galactic globular clusters~\citep{2012Natur.490...71S,2013ApJ...777...69C,2015MNRAS.453.3918M}.
In fact, if not a tMSP, the aforementioned UCXB 47 Tuc X9 most likely has a BH accretor \citep{Bahramian:2017mbs}.
We will thus assume that also UCXBs with BH accretors exist. 
Given that (1) about one dozen 
of UCXBs are detected of which one is potentially a BH-UCXB
\citep{vanHaaften:2012qc,Bahramian:2017mbs}
and 
(2) the numbers of detected bright LMXBs
compared to the number of detected LMXBs with BH accretors 
is consistent with about $\sim30\%$ being BH-LMXBs
\citep{2007A&A...469..807L, Tetarenko:2015vrn,Miller-Jones:2015zba},
a fraction of $\mathcal{O}(10\%)$ UCXBs
with BH accretors is not unrealistic.

To summarise, we will make the following assumptions:
\begin{enumerate}
  \item{A population of NS-UCXBs is present in the bulge
  \citep{vanHaaften:2013uuo}, providing the
  MSPs required for the GCE.}
  \item{An analogous population of BH-UCXBs is present, which
  we assume can make up a $\mathcal{O}(10\%)$ fraction of the total
  population.}
  \item{
  Due to their low mass-accretion rates today, most UCXBs 
  are in the low/hard or quiescent state and 
  can thus emit positrons as 
  described in Sect.~\ref{sec:bin_pos}. 
  We model the NS-UCXBs according to the radio--X-ray correlation observed for tMSPs (see Sect.~\ref{sec:NS}) in
  which case they can have positron-injecting jets about 20\% of the time. We deem this as a reasonable assumption since tMSPs are the most
  quiescent NS-LMXBs currently present in the radio--X-ray correlation and because the UCXB Tuc 47 X9 lies on the same radio--X-ray 
  correlation as the other tMSPs.} 
\end{enumerate}

\subsubsection{Modeling of NS- and BH-UCXBs}
Next, we discuss how we model the population of NS- and BH-UCXBs in
the Galactic bulge and their positron yield.
The discussion builds upon the work by \cite{vanHaaften:2011iy, vanHaaften:2013uuo}.
Star formation in the bulge is taken to happen early on during 
its lifetime
and is modeled with a narrow gaussian
centered $10\mathrm{\,Gyr}$ ago with a standard deviation of 
$0.5\mathrm{\,Gyr}$: 
$T_\mathrm{sf} \sim \mathcal{N}\left(\mu=-10\mathrm{\,Gyr}, \sigma = 0.5\mathrm{\,Gyr}\right)$.
For a total bulge mass of $10^{10}\mathrm{\,M_\odot}$, we assume a yield
of $N_\mathrm{NS-UCXB}=2\times 10^{5}$ 
NS-UCXBs, as suggested by the results of 
\cite{vanHaaften:2013uuo}. 
The majority of these NS-UCXBs have either a helium star or WD donor.
Roche-lobe overflow occurs faster for Helium star donors and they will
therefore have a shorter delay time between zero-age main 
sequence\footnote{The time at which a star 
first enters the main sequence, in this
case the time of star formation.}
(ZAMS) and the start of mass transfer.
The delay-time (DT) distributions, for the delay between ZAMS
and the start of Roche-lobe overflow of He-star and WD donors
onto a NS, are given in Fig.~3 of
\cite{vanHaaften:2013uuo}. 
He-star donors typically have much shorter delay times
than WDs, with a median (mean) of $\sim 0.1 \,(0.2)\mathrm{\,Gyr}$,
compared to $\sim 1.5 \,(2.2)\mathrm{\,Gyr}$ for WDs.
For the purpose of our paper,
we will assume that the delay-time
distribution and the distribution of donors for BH-UCXBs are
\textit{identical} to that of NS-UCXBs, the only difference
being an overall rescaling factor for the ratio of BH-to-NS systems. 

\begin{figure}
	\includegraphics[width=\columnwidth]{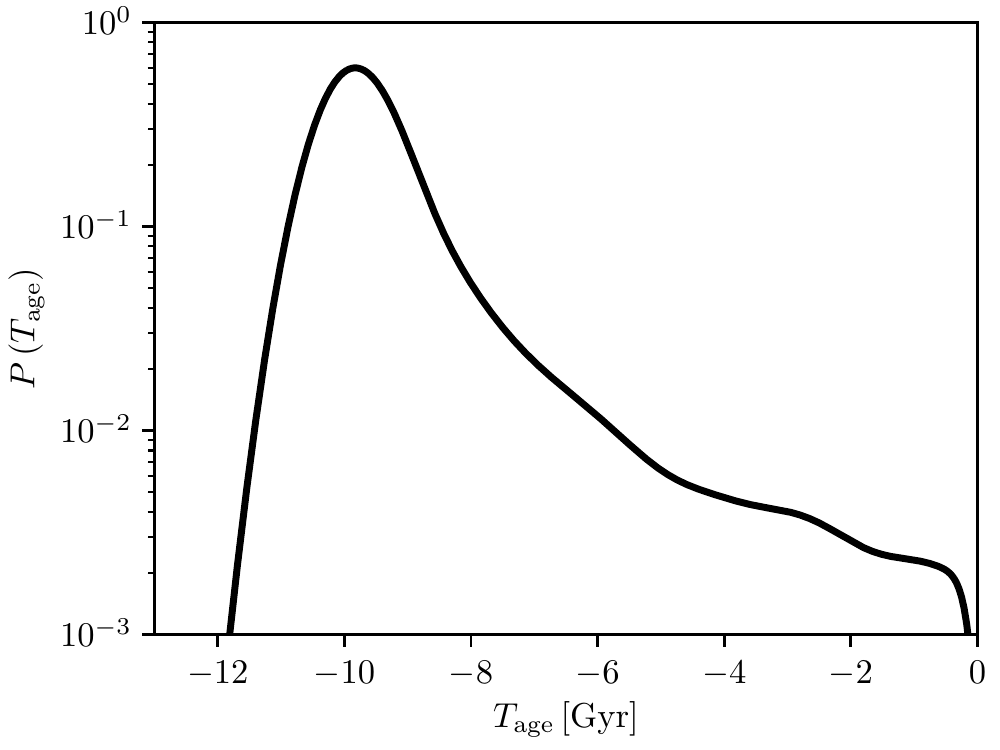}
    \caption{
	Age distribution of the UCXBs in our analysis normalized
    to unity. 
    $T_\mathrm{age}$ refers to the time at which 
    the compact object started accreting from its donor,
    with $T_\mathrm{age}=0$ referring to today.
    }
    \label{fig:age}
\end{figure}

We refer
to the delay-time by $T_\mathrm{DT}$. 
Using this we can determine the age of the
UCXB through
\begin{equation}
	T_\mathrm{age} = \left(T_\mathrm{sf} + T_\mathrm{DT}\right).
\end{equation}
Note that $T_\mathrm{age}$ is defined as a negative number.
Typical ages are for UCXBs with He-star (WD) donors
are $\sim 10\, (8)\mathrm{\,Gyr}$.
We then assume that the probability for some delay-time is 
independent of the time of star formation. This allows us to easily
find the distribution of $T_\mathrm{age}$, which peaks
around $-10\mathrm{\,Gyr}$ and has a tail until $T_\mathrm{age}=0$
(today) due to the delay-time distribution being skewed towards large delay times (see Fig.~\ref{fig:age}).

The evolution of UCXBs was studied in \cite{vanHaaften:2011iy}.
They provide both fitted tracks and analytic approximations 
for the mass-accretion rate of a UCXB with a given age. 
The result of the analytic approximation is
\begin{align}
	\label{eq:Mdot}
	\dot M &\approx 2.75 \times 10^{-12}
    \left(\frac{M_\mathrm{a}}{\mathrm{M_\odot}}\right)^{-\frac{2}{11}}
    \left|\frac{T_\mathrm{age}}{1\mathrm{\,Gyr}}\right|^{-\frac{14}{11}} 
    \mathrm{\,M_\odot\,yr^{-1}},
\end{align}
with $M_\mathrm{a}$ the mass of the accreting object. We will
take $10\,(1.4)\mathrm{\,M_\odot}$ for a BH (NS). 
Due to the small exponent the overall results are not very sensitive
to the exact value of the mass. What should be noted though is
that for a given age, BHs typically accrete less quickly than NSs.

Having obtained the accretion rates we can proceed
to calculate the X-ray luminosity,
jet power, positron yield and 511 keV emission using
Eqs (\ref{eq:Lbol}\text{--}\ref{eq:N511}).
\begin{figure}
	\includegraphics[width=\columnwidth]{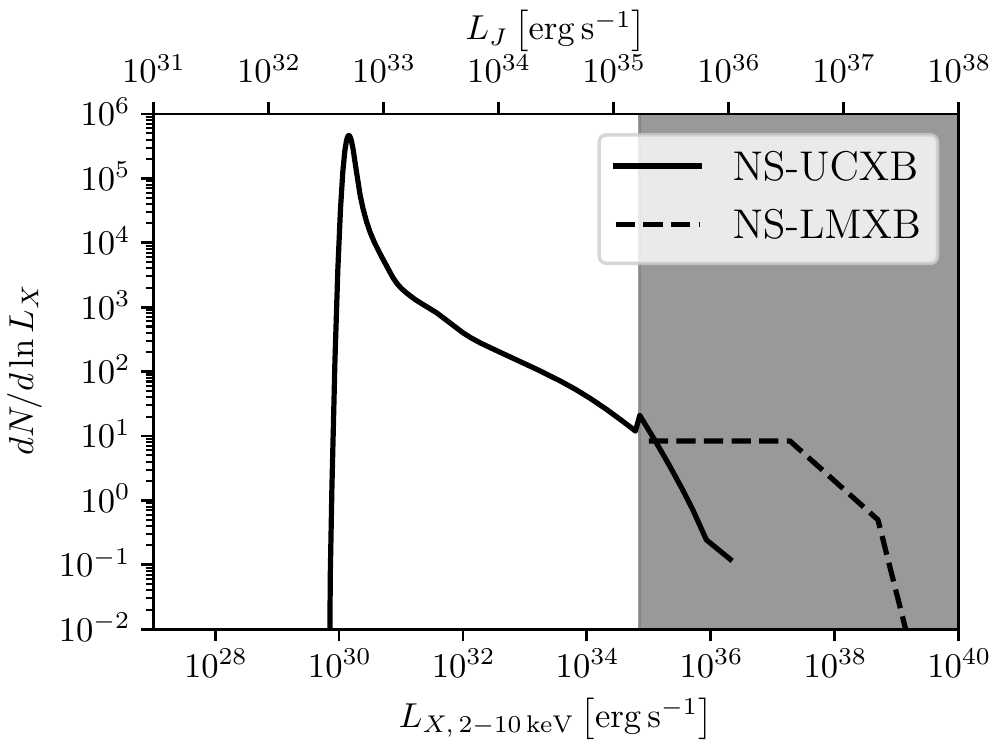}
    \caption{
    The luminosity function of NS-UCXBs (solid), using
    $N_\mathrm{NS}=2\times10^5$.
     There is a pile-up of old
    sources with low X-ray luminosities.
    For reference we also show the distribution of bright LMXBs (dashed) from
    \citet{Gilfanov:2003th}.
    Dim X--ray sources can still have considerable 
    jet power and thus a sizeable positron outflow. 
    The grey-shaded area corresponds to
    $\dot M> \dot M_\mathrm{crit}$ (1\% of the Eddington
    accretion rate) where the jet dominated state can be expected
    to terminate. The transition from the low/hard to high/soft state also causes the kink in the solid line.
    }
    \label{fig:Lx_NS}
\end{figure}
\begin{figure}
	\includegraphics[width=\columnwidth]{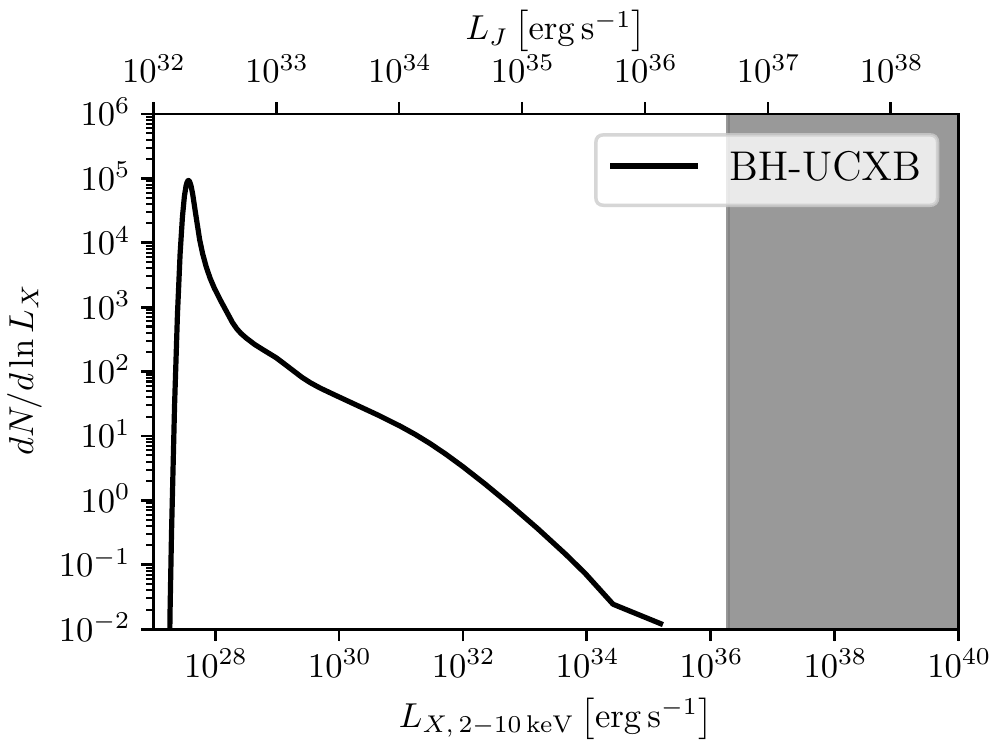}
    \caption{
    Similar to Fig.~\ref{fig:Lx_NS} but for BH-UCXBs ($N_\mathrm{BH}=4\times10^4$). 
    }
    \label{fig:Lx_BH}
\end{figure}
The distribution of NS-UCXBs and BH-UCXBs with a given X-ray power and given jet power
are shown in Figs~\ref{fig:Lx_NS} and \ref{fig:Lx_BH}  (solid lines). 
All BH-UCXBs should be in a low/hard or quiescent
state, as can be seen from Fig.~\ref{fig:Lx_BH} where
the grey-shaded area corresponds to 
$\dot M> \dot M_\mathrm{crit}$. Similarly, most of the NS-UCXBs will
be in the low/hard state. We assume these systems have jets.
There is a large pile-up of systems at low mass-accretion rates/X-ray brightness, a
consequence of most UCXBs being old systems (see Eq.~\ref{eq:Mdot})
\citep{vanHaaften:2013uuo}. The integrated X-ray luminosity of this
population is still dominated by the brightest sources,
however, the total jet power, and therefore the total number of
positrons injected is dominated by the dimmest sources. This can
be understood from the dependence on the accretion rate
which is $L_X \propto \dot M^2$ whereas $L_J \propto \dot M$.
In Fig.~\ref{fig:Lx_NS} we also show for reference the distribution of 
bright LMXBs from \cite{Gilfanov:2003th}. Since this distribution fully lies
in the grey-shaded area, meaning these sources are not inefficiently accreting, we do not expect steady jets from them.

\begin{figure}
	\includegraphics[width=\columnwidth]{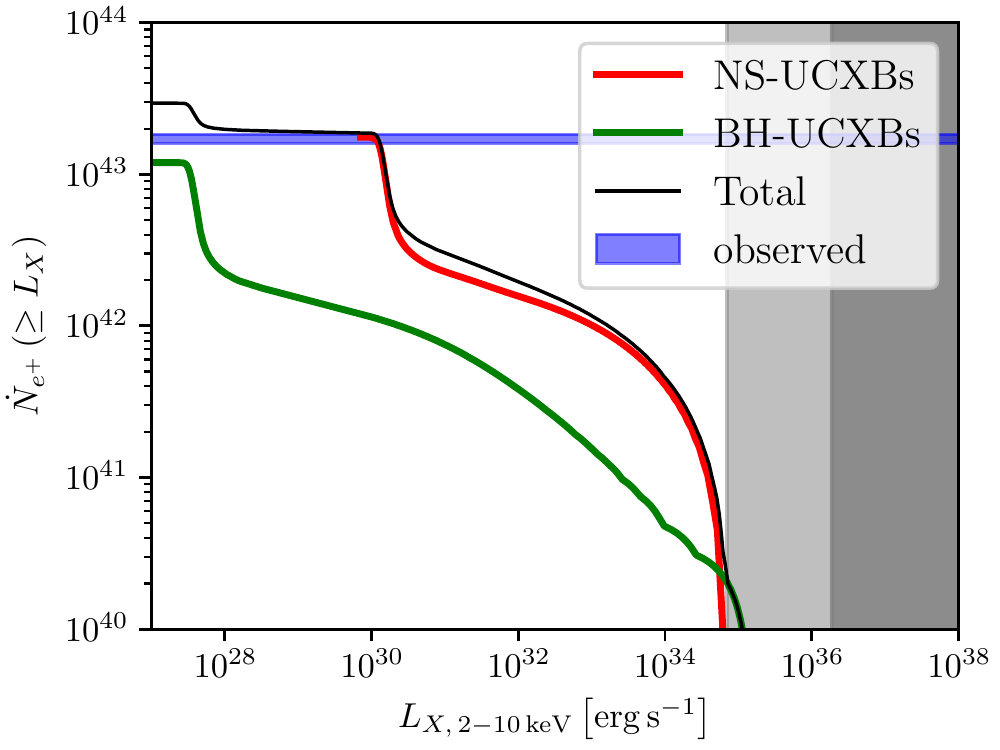}
    \caption{
	Total number of positrons injected into the ISM by
    NS-UCXBs/BH-UCXBs above a given X-ray luminosity (red/green) $L_{X,\,2-10\mathrm{\,keV}}$. The black
    solid line shows the total cumulative amount of injected
    positrons from all UCXBs.
    Assumed jet properties are: $\Gamma = 1.4$,
    $\left<\gamma\right>=1$, $A_\mathrm{BH}=0.1$
    and $A_\mathrm{NS}=0.03$. The total number
    of NS-UCXBs (BH-UCXBs) is 
    $2\times10^{5}$ ($4\times10^{4}$).
    Due to their large numbers 
    the quiescent X-ray binaries are the dominant source of positrons.
    The blue band shows the observed positrons injection rate in the bulge from \citep{Siegert:2015knp} assuming a distance
    to the bulge of 8.3 kpc.
    }
    \label{fig:positrons}
\end{figure}

In Fig.~\ref{fig:positrons} we show the cumulative positron 
yield from the population of UCXBs in the Galactic bulge.
The injection of positrons is dominated by the quiescent 
sources due to their large numbers (see Figs~\ref{fig:Lx_NS} and \ref{fig:Lx_BH}).
The $2\times10^5$ NS-UCXBs by themselves can already reproduce the observed injection rate of positrons,
$2\times 10^{43}\mathrm{\,s^{-1}}$ \citep[e.g.~][]{Knodlseder:2005yq,
Siegert:2015knp} in the Galactic bulge,
when their jets are modeled as described in Sect.~\ref{sec:bin_pos}.
The BH-UCXBs inject slightly fewer positrons
in our model than their NS counterparts.
This is explained by our assumption that they have
an identical evolutionary channel as the NS-UCXBs.
Consequently, they will have lower mass-accretion rates
for a given age due to their higher mass (see Eq.~\ref{eq:Mdot}). Their overall smaller number 
is counterbalanced by the fact that we assumed a
20\% duty cycle for the NS-UCXBs.

\section{Future avenues}
\label{sec:prospects}
In this section we present a number of directions that can be pursued in the 
(near) future to further study the hypothesis that the 511 keV signal comes
predominantly from X-ray binaries and/or that the GCE and positron-annihilation signal
are indeed connected.
\subsection{Comparing morphologies}
In Sect.~\ref{sec:intro} we argued that the morphology of the GCE and bulge 
511 keV signal appear to be similar. However, this is, to a large extent, only a qualitative statement.
Detailed spectroscopic analyses of the 511
keV line suggest that most annihilation takes place in a warm-ionized and
warm-neutral medium \citep{Churazov:2004as, Guessoum2005, Jean:2005af}.  This
scenario is in good agreement with a diffuse distribution of sources that
inject low-energy ($\lesssim 1\mathrm{\,MeV}$) positrons.  Such positrons are
unlikely to escape the warm medium and will annihilate within $\sim50\mathrm{\,pc}$
of their injection sites \citep{Jean:2005af}. Consequently, it is likely that
the positron-annihilation signal follows closely the source distribution.
Therefore, in the case of a related origin to the GCE, it is also expected to closely
trace the GCE.

\paragraph*{511 keV model-fitting}
Although SPI has imaging capabilities, the most accurate analysis results
are obtained with model-fitting \citep[e.g.~][]{Strong:2005zx}. For instance, 
the latest results from \cite{Siegert:2015knp} use a model consisting of a superposition
of gaussians representing a disk,
narrow and broad bulge, and a Galactic Center component. 
On the other hand, \cite{Vincent:2012an} use an Einasto profile in their analysis
and they argue that it also provides a good fit. The Einasto profile is more peaked towards
the Galactic center than the superposition of gaussians, even including the Galactic center component.
Unfortunately, the combination of SPI its broad point-spread function and limited exposure is insufficient to distinguish between these
profiles. 
In the context of the \Fermi-LAT GCE, \cite{Bartels:2017vsx} showed that using the superposition of two 
gaussians plus a central source from \cite{Siegert:2015knp} to model the GCE performs
worse than a template tracing the
distribution of stars in the boxy bulge, but slightly better
than DM inspired templates.
A next step
towards a quantitative comparison of the two emissions would be to consider a physically motivated bulge
template in the analysis of the 511 keV emission, similar to the one adopted in \cite{Bartels:2017vsx}, in a model-fitting analysis
of the 511 keV line.

We also point out that the intensity of the 511 keV emission is consistent with
tracing the
distribution of stellar mass in the inner-Galaxy. 
In \cite{Bartels:2017vsx} we argued that the GCE is well traced by the distribution
of stellar mass in the Galactic bulge. 
A template consisting of a combination of nuclear bulge \citep{Launhardt:2002tx} and the
boxy bulge \citep{Cao:2013dwa} scaled according to their relative masses provides
a good fit to the GCE data \cite{Bartels:2017vsx}.
If the GCE traces stellar mass, and if the GCE is connected to the 511 keV signal,
we expect that the 511 keV signal traces stellar mass. 
INTEGRAL-SPI would see the nuclear bulge as only marginally extended and  difficult to discriminate from a point source. 
\cite{PoS(INTEGRAL2014)054} and \cite{Siegert:2015knp} found evidence for the presence of a central Galactic source
(GCS), on top of their gaussian bulge model. This source is compatible with being point-like and has a maximum extent of a few-hundred parsec, consistent with the nuclear bulge.
The fluxes from the central source and bulge are
$(0.8\pm0.2)\times 10^{-4}\mathrm{\,ph\,cm^{-2}\,s^{-1}}$ and 
$(9.6\pm0.7)\times 10^{-4}\mathrm{\,ph\,cm^{-2}\,s^{-1}}$, respectively. The mass of the nuclear bulge is
estimated at $M_\mathrm{NB} = 1.4\pm 0.6\times 10^9\mathrm{\,M_\odot}$
\citep{Launhardt:2002tx}.
Scaling the mass of the 
nuclear bulge to that of the larger Galactic bulge by 
the ratio of 511 keV emission from the GCS and the bulge
at large, we find
$M_\mathrm{B} = 1.7\pm 0.8\times 10^{10}\mathrm{\,M_\odot}$, in agreement
with mass estimates in the literature which are of the order 
$\sim 1\times 10^{10}\mathrm{\, M_\odot}$ \citep{Licquia:2014rsa,Cao:2013dwa}. Therefore, the 511 keV distribution seems roughly consistent
with the distribution of stellar mass in the bulge.

\paragraph*{Asymmetry in the 511 keV signal}
\cite{Weidenspointner:2008zz} reported an asymmetry in the 511 keV emission. This 
asymmetry was ascribed to an asymmetry in the disk, with the west ($\ell<0$) being
brighter. 
Notably, the authors related the 511
keV asymmetry to a similar asymmetry in the distribution
of LMXBs. 
However, later analyses do not support
the conclusion of an asymmetry in the disk \citep{Bouchet:2010dj,Siegert:2015knp}. 
Rather, these analyses find a preference for a slightly shifted central component towards
negative longitudes. 
On the other hand, it should be noted that no such asymmetry is observed in the GCE
\citep{Bartels:2017vsx}, 
and that a boxy bulge model is in fact brighter at positive longitudes (east), 
due to its rotation
\citep[e.g.~][]{Cao:2013dwa}.
We argue that this apparent discrepancy in the context of a potential correlation between the GCE and
511 keV signal is not of major concern at this stage. First, it is important to compare in
more detail the signal morphologies, as suggested above. Secondly, if present, some asymmetry towards
negative longitudes might be due to the presence of the Sagittarius-Carina spiral arm along the 
line-of-sight \citep{Alexis:2014rba}.

\paragraph*{Bulge-to-disk ratio}
\cite{Siegert:2015knp} detect the disk in 511 keV line emission, with a bulge-to-disk flux ratio 
of about 0.6. Currently, no such disk
has revealed itself in GeV $\gamma$--rays \citep{Bartels:2017vsx}. In 
the scenario of \cite{Crocker:2016zzt} and also in the scenario presented above some disk component
of the GCE would be expected due to the foreground MSPs present in the disk. Currently,
uncertainties in GeV $\gamma$-rays are too large to claim any detection of a disk. However, at the moment
the B/D flux-ratio in GeV $\gamma$-rays is
consistent with $\sim 0.5$.
Furthermore, the bulge-to-disk ratios of the GCE and 511 keV signal do not need to be identical.
For instance, a sizeable contribution of positrons from radiaoctive isotopes produced in
stellar nucleosynthesis is expected in the Galactic disk, but not necessarily in the bulge.
This could lead to a brighter disk in 511 keV compared to its GCE counterpart.

\subsection{511 keV imaging with IBIS}
The imager on board the INTEGRAL satellite (IBIS) has poorer spectral resolution 
($\Delta E / E \sim 0.1$) than SPI, but has a much better spatial resolution:
$\sim 12'$ ($\sim 2.5^\circ$ for SPI) \citep{Ubertini:2003ih}.
It has been used to search for 511 keV point sources by \cite{DeCesare:2011gc}.
The energy window of interest used in that work is ($491\text{--}531\mathrm{\,keV}$). 
In this window, the emission from positronium annihilation
in a large part of the inner-Galaxy is so bright that it
outshines the continuum background. 
Note that since the 511 keV line is almost entirely due to 
positronium annihilation instead of direct annihilation, i.e.~the
positronium fraction is close to 1, the majority of the photons in
the aforementioned energy window are from the 3-photon o-Ps final state and produce
a continuum rather than a line, in a ratio
2:9 photons from p-Ps versus o-Ps. However, since 
both trace
positronium directly, we are interested in their sum.
From the results of \cite{Siegert:2015knp} 
we estimate that over $90\%$ of the photons from the inner
$40^\circ\times40^\circ$ should trace the distribution of Ps 
(for details see Appendix \ref{sec:IBIS}).
Therefore, it should in principle be possible to 
use the imaging capabilities of IBIS to study in more detail the morphology of the 511
keV signal in the bulge.

However, at energies $>100\mathrm{\,keV}$ IBIS
has a large cosmic-ray-induced internal background which is very high, $\sim 10\mathrm{\,ph\,s^{-1}}$, in the energy window of interest \citep{Lebrun:2003aa, DeCesare:2011gc}. 
Consequently, any signal is swamped by the instrumental backgrounds. An understanding of this background below the percent level is required to study diffuse emission features. 
Alternatively, if the diffuse 511 keV signal is comprised of many dim point sources, one can use a wavelet-based approach to filter out the backgrounds, similar to what was done by \cite{Bartels:2015aea} to study the origin of the Galactic-Center excess in the \Fermi-LAT data.
In appendix \ref{sec:IBIS} we show, under very simple assumptions, that such an analysis on IBIS data 
can in principle  reveal a point source origin of the 511 keV emission.
This exercise represents a first proof of principle that the method can potentially work, 
however, a more detailed and involved modeling of instrumental effects and coded-mask performance would be required in order 
to assess the sensitivity of such a search. 
Such a simulation is beyond the scope of this work.

\subsection{511 keV emission from Galactic globular clusters}
\label{sec:MWGC}
Globular clusters (GCs) are known to host MSPs and LMXBs.
$\gamma$--ray emission is detected from GCs, and it
most likely originates from their MSPs~\citep{Zhang:2016tcf}.
The presence of $\gamma$-rays from MSPs and the presence
of LMXBs suggests, according to our scenario, that a 511 keV line 
emission could also be present.
Indeed, GCs can also harbour a population of UCXBs,
and the discovery of the first potential BH-UCXB in the Galactic globular cluster
Tucanae 47 \cite{Bahramian:2017mbs} could be just the tip of the iceberg.

We investigate here the sensitivity of INTEGRAL/SPI to the detection of 511 keV line emission
from GCs in the Milky Way (MWGCs). We select the 16 globular clusters detected
in $\gamma$-rays with \Fermi-LAT and quoted
in~\citet{Zhang:2016tcf}. We estimate what their 511 keV line emission should be
using the ratio of 511 keV to GCE emission observed in the bulge.
The integrated GCE flux above $0.1\mathrm{\,GeV}$ is 
$\left(1.6\pm0.1\right)\times 10^{-6}
\mathrm{\,GeV\,cm^{-2}\,s^{-1}}$ 
\citep{Bartels:2017vsx} and the 511 keV flux is
$(10.4\pm0.7)\times 10^{-4}\mathrm{\,ph\,cm^{-2}\,s^{-1}}$ 
\citep[][for the bulge and central source]{Siegert:2015knp}.
So the number of 511 keV photons per GeV is:
$(1.5\pm0.1)\times10^{-3} \mathrm{\,GeV\,ph^{-1}}$. 
Knowing the exposure\footnote{
Available from \url{https://www.cosmos.esa.int/web/integral/exposure-map-tool}.
The exposure shows large
fluctuations on the sky \citep[][Fig.~1]{Siegert:2015knp}. In the inner
Galaxy it is $\sim 2\times10^{9}\mathrm{\,cm^2\,s}$, but it can be smaller
by a few orders-of-magnitude at other positions on the sky.}
in the direction of the selected globular clusters allows us to test
whether a specific MWGC can be observable by SPI.
The SPI effective area at 511 keV is $\sim 75\mathrm{\,cm^2}$
and the 2$\sigma$ narrow-line sensitivity is 
$5.7\times10^{-5} \sqrt{10^{6}/T_\mathrm{obs}}
\mathrm{\,ph\,cm^{-2}\,s^{-1}}$ \citep{Siegert:2016ijv}.

\begin{figure}
	\includegraphics[width=\columnwidth]{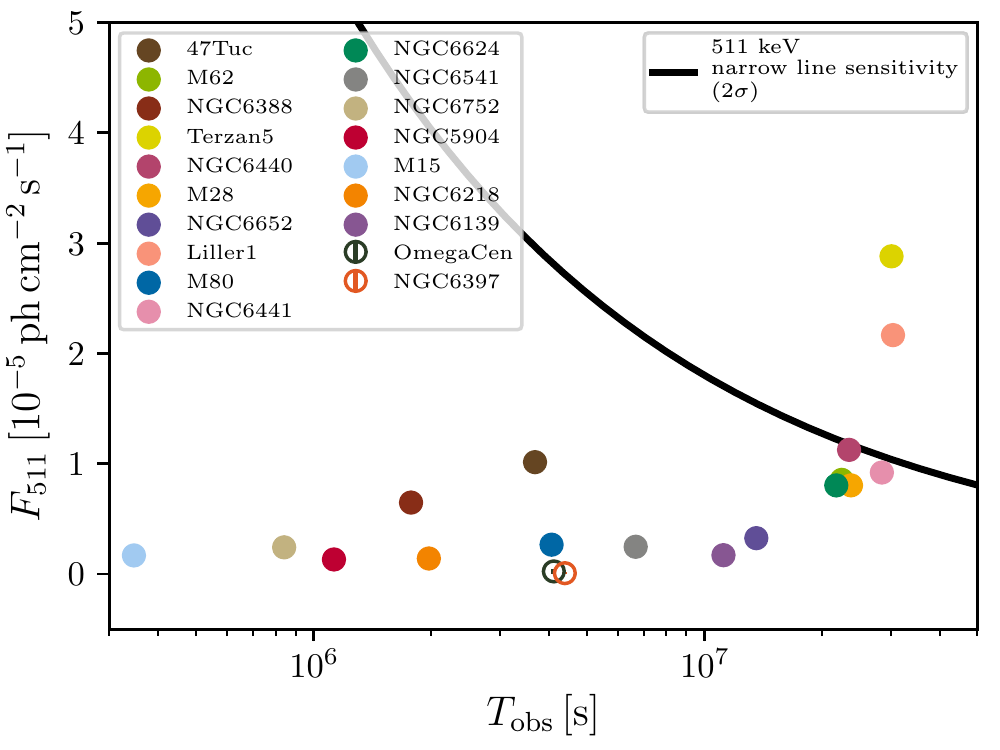}
    \caption{
    Sensitivity to MWGCs observed in $\gamma$-rays.
    Colored circles give the 511 keV flux of 
    $\gamma$-ray detected GCs 
    assuming that the
    number of 511 keV photons scales 
    with the MSPs $\gamma$-ray luminosity as observed
    in the bulge: $1.5\times 10^{-3}\mathrm{\,GeV/ph}$. 
    On the horizontal axis we show the total SPI exposure time
    at the source position.
    The solid line
    gives the SPI sensitivity to a narrow line for a given
    observation time \citep{Siegert:2016ijv}.
    Upper limits exist in the literature for some GCs
    \citep[open circles][]{Knodlseder:2005yq}.
    }
    \label{fig:MWGCs}
\end{figure}

In Fig.~\ref{fig:MWGCs} we show the sensitivity to a narrow 511 keV
line against the exposure time at a given source position 
(black solid). 
Colored circles show the 511 keV flux estimated from the $\gamma$--ray
luminosity of the source. For open circles an upper limit
already exists in the literature \citep{Knodlseder:2005yq}, 
all about a factor of $\sim 5$ above the current sensitivity curve.
According to this estimate, 3 MWGCs (Terzan 5, Liller 1 and NGC 6440) 
can potentially be seen in the INTEGRAL/SPI data. 
The estimated fluxes of another 4 MWGCs
(M62, M28, NGC 6441 and NGC 6624) lie within a factor 2
of the narrow-line sensitivity.
We note that all of these sources lie within the bulge region
and would constitute $\mathcal{O}(10\%)$ of the
the total 511 keV bulge emission in the above estimate.
Including the 
GCs as separate sources would absorb part of the bulge flux,
and thus translate into a lower estimated number of 511 keV
photons per GeV from the GCE and thus a lower predicted GC flux.
Again, we stress that an analysis of the 511 keV line emission 
with IBIS can play a crucial role in this context. 

Taking the 2$\sigma$ sensitivity flux as an estimate of an upper
limit on the 511 keV narrow-line emission for each GC, we
estimate an ``upper limit" on the stacked 511 keV emission from $\gamma$--ray
GCs,  
$\sqrt{\sum_i (UL)_i^2} \sim 1.3 \times 10^{-4}\mathrm{\,ph\,cm^{-2}\,s^{-1}}$.
This value is in the ballpark, slightly below, the 511 
keV flux expected from stacking the $\gamma$--ray
emission of the 16 globular clusters ($2.1 \times 10^{-7}\mathrm{\,GeV\,cm^{-2}\,s^{-1}}$) and
computed assuming the conversion mentioned above, i.e.~
$1.5\times 10^{-3}\mathrm{\,GeV\,ph^{-1}}$.

The above estimate is valid under the assumptions that (a) the LMXB
to MSPs ratio in GCs today is comparable with the one in the Galactic bulge --
which might be the case in a scenario where the Galactic bulge stellar
population is created by disruption of old GCs --,
and (b) there is enough gas for
the positrons to annihilate before leaving the cluster 
or in the direct vicinity of the cluster.

Observing $511\mathrm{\,keV}$ emission from Galactic GCs would be a major step in
claiming a connection between the GCE and low-energy positrons and in 
addition would make X-ray binaries the primary candidate to produce most of the
Galactic positrons. 
In addition, it can provide insights into the connection between the Galactic
bulge and the GCs.
However, we emphasize that evolutionary channels in the bulge and in GCs need not be 
the same.
For instance, dynamical interactions are more likely in GCs due to the high stellar density
\citep{Ivanova:2007bu}. 
As such, a non-observation would not be able to rule out a 
population synthesis scenario. 
Nevertheless, an analysis of GCs similar to that performed by \cite{Siegert:2016ijv}
for dwarf spheroidals can shed more light on a potential
GCE-511 keV connection.

\subsection{Individual source searches}
The direct observation of a dim X-ray binary with a 
steady-state jet producing positrons would identify
these sources as a prime candidate to produce the majority of
the Galactic positrons. Quiescent sources in the vicinity of the sun
are the most promising candidates since they would have the highest
observable flux. However, we note that the
flux and the spatial extend of the signal also depend on the
ISM in the vicinity of the source.
In particular the transitional millisecond pulsars
J1023+0038 \citep[e.g.~][]{Deller:2014yca} and XSS J12270-4859
\citep{Bassa:2014hoa, Roy:2014cwa} are promising targets.
Both sources are listed as $\gamma$-ray sources associated with pulsars
in respectively the 2FGL and 3FGL \citep{Fermi-LAT:2011yjw, Acero:2015hja} and as $\gamma$-ray detected MSPs
in the preliminary \Fermi-LAT 8-year source catalog
\footnote{The FL8Y catalog is available from \url{https://fermi.gsfc.nasa.gov/ssc/data/access/lat/fl8y/}. Also see
\url{https://confluence.slac.stanford.edu/display/GLAMCOG/Public+List+of+LAT-Detected+Gamma-Ray+Pulsars} for the public list of $\gamma$-ray detected pulsars.}
(FL8Y).
Moreover, since both sources are at distances
$\sim 1.4\mathrm{\,kpc}$ their 511 keV flux is potentially already
detectable by INTEGRAL/SPI. A discovery of $\gamma$-rays from these
sources would be a smoking gun for a one-to-one relation between
the 511 keV signal and the GCE. These tMSPs and
a few promising BHs candidates are discussed in some
more detail in Sect.~\ref{sec:individual_sources}.
In upcoming years, proposed next-generation $\gamma$-ray telescopes
such as e-ASTROGAM, which has about one order-of-magnitude better
sensitivity to a 511 keV narrow line than SPI \citep{DeAngelis:2016slk, DeAngelis:2017gra}, can greatly improve the sensitivity in searches for such individual sources.


\section{Discussion \& conclusion}
\label{sec:discussion}

\paragraph*{511 keV emission and LMXBs.}
In this paper we proposed a model in which LMXBs, in particular a subclass called
ultracompact X-ray binaries, are responsible for the positron annihilation signal in the
bulge. Depending on assumptions, either NS-LMXBs
or BH-LMXBs can be responsible for the majority of
the positrons. In our specific modeling, where
we adopted the evolutionary model of 
\cite{vanHaaften:2011iy,vanHaaften:2013uuo} with $\sim 2\times 10^5$
NS-UCXBs,
NS-UCXBs produced most positrons. 
We here assumed NS-UCXBs
jets to behave similarly to what is expected for
tMSPs \citep{Deller:2014yca}.
In this case
a single source class, NS-UCXBs, is able to produce both the GCE and 511 keV emission. 

The binary jets are assumed 
to be loaded with non-relativistic electron-positron pairs 
\citep[][also see Sect.~\ref{sec:bin_pos}]{Heinz:2002qj}.
Whether this is an accurate model for these systems in the quiescent or low/hard state
remains to be seen. 
However, more freedom on the jet modeling is allowed if more X-ray binaries
are present in the bulge. 
\cite{Tetarenko:2016uln}
recently observed a quiescent BH-LMXB in the field. 
Taking into the consideration the surveyed area in their analysis, 
they extrapolate the observation of this single source 
to a total Galactic population of $2.6\times10^4$--$1.7\times10^8$ of these objects, the latter number being
vastly greater than the $\sim 2\times10^{5}$ objects 
we assumed.

\cite{Bandyopadhyay:2008ts} also argued that quiescent LMXBs could produce all the positrons
in the Galactic bulge. A further similarity between their analysis and this one is that both assume the presence of electron-positron
pairs in the jet. However, the average assumed jet power in their analysis is higher 
than the average of our population, $10^{35}\mathrm{\,erg\,s^{-1}}$ vs. 
$\sim 10^{33}\mathrm{\,erg\,s^{-1}}$, consequently, our population has to be larger. 
In addition, 
we have shown explicitly 
that the accreting object need not
necessarily be a BH, but can also be a NS.
The NS contribution was considered subdominant by \cite{Bandyopadhyay:2008ts}.
Alternatively, positrons could also come from brighter X-ray binaries, called microquasars, as proposed
by \cite{Guessoum:2006fs}. Observational evidence for production of positrons in these sources comes comes from \cite{Siegert:2016ymf}. However, \cite{Bandyopadhyay:2008ts} argue that even
in the presence of microquasars, quiescent sources would dominate the total positron injection rate
in their scenario. Moreover, if all positrons are injected during bright outburst of which only
a handful take place at any given time we should at some stage start resolving the individual sources.

At this point in time, the scenario outlined in this paper is realistic in terms of our
theoretical and observational understanding of inefficiently accreting Galactic binaries, both in terms of their number and
in terms of their jet composition. Consequently, the family tree of MSPs, the sources assumed responsible for the GCE, could also host the sources of Galactic positrons. We stress again that
the specific model is not the only possible model for this scenario. Rather, it serves as a 
prototype for the viability of the GCE--511 keV connection.

\paragraph*{Arguments against an MSP interpretation of the GCE.}
It has been argued that MSPs cannot explain the GCE
because of the lack of bright LMXBs observed in the
Galactic bulge when compared to GCs. More bright LMXBs would 
be expected based on the relative $\gamma$-ray brightness
of the bulge and GCs
\citep[e.g.~][]{Hooper:2013nhl, Cholis:2014lta, Haggard:2017lyq}.
We point out that the UCXBs in this scenario
are very dim in X--rays,
as can be seen in Figs~\ref{fig:Lx_NS} and \ref{fig:Lx_BH}. 
Therefore, while producing enough MSPs to explain
the GCE, the number of X-ray sources is also consistent with observation \citep{Jonker:2011gf, vanHaaften:2013uuo, vanHaaften:2015wta, Hong:2016qjq}.
In this scenario, the MSPs and UCXBs are directly produced in the bulge and not 
for instance associated
with the disruption of globular cluster \citep[e.g.][]{Brandt:2015ula}.
Thus, an MSP interpretation of the GCE
is perfectly compatible with the different ratios of LMXBs versus GeV emission 
in GCs and in the Milky-Way bulge. 
Arguments against an MSP interpretation based on the 
non-observation of LMXBs are fully dependent on the assumed formation channel of MSPs
\citep[of which there are many,][]{Ivanova:2007bu} and the assumption
that populations in both GCs and the bulge are similar. 
A similar statement is made by \cite{Ploeg:2017vai}.

\paragraph*{Closing remarks.}
The GCE and 511 keV emission share striking morphological 
similarities making a common scenario that can explain both at the same time appealing.
Inspired by the literature \citep{vanHaaften:2013uuo}, we have presented a population synthesis model for UCXBs in the Galactic bulge which can account for the required number of MSPs to explain the GCE and, with some additional assumptions,
can also be at the origin of the low-energy positrons through
jets.
We can conclude that both NS-UCXBs and
BH-UCXBs could sustain and be the dominant contributor
to the 511 keV flux depending on how exactly they are modeled.
To the best of our knowledge this is the first work that quantitatively shows NS-LMXBs can sustain the positron injection rate. In fact, if these NS binaries in the low-hard state behave like tMSPs,
the $\gamma$-ray sources producing the GCE could themselves be injecting the low-energy positrons, thus a single source class can explain both.
The model presented
in this works serves as a prototype for explaining a common origin 
of the GCE and the 511 keV emission and offers 
new windows to study their connection.


\section*{Acknowledgements}

We would like to especially thank Riley Connors, Gijs Nelemans
and Thomas Siegert for discussion and comments. We also thank
Roland Crocker, Torsten En{\ss}lin, Amruta Jaodand, Fiona Panther
and Volodymyr Savchenko for useful discussion.
This research is funded by NWO through the VIDI research program
"Probing the Genesis of Dark Matter" (680-47-532; ES, CW) and through a GRAPPA-PhD program (022.004.017; RB).
FC acknowledges support from Agence
Nationale de la Recherche under the contract ANR-15-IDEX-02, project ``Unveiling the Galactic centre mistery",
GCEM (PI: F. Calore).




\bibliographystyle{mnras}
\bibliography{GCE_511}

\begin{thebibliography}{}
\makeatletter
\relax
\def\mn@urlcharsother{\let\do\@makeother \do\$\do\&\do\#\do\^\do\_\do\%\do\~}
\def\mn@doi{\begingroup\mn@urlcharsother \@ifnextchar [ {\mn@doi@}
  {\mn@doi@[]}}
\def\mn@doi@[#1]#2{\def\@tempa{#1}\ifx\@tempa\@empty \href
  {http://dx.doi.org/#2} {doi:#2}\else \href {http://dx.doi.org/#2} {#1}\fi
  \endgroup}
\def\mn@eprint#1#2{\mn@eprint@#1:#2::\@nil}
\def\mn@eprint@arXiv#1{\href {http://arxiv.org/abs/#1} {{\tt arXiv:#1}}}
\def\mn@eprint@dblp#1{\href {http://dblp.uni-trier.de/rec/bibtex/#1.xml}
  {dblp:#1}}
\def\mn@eprint@#1:#2:#3:#4\@nil{\def\@tempa {#1}\def\@tempb {#2}\def\@tempc
  {#3}\ifx \@tempc \@empty \let \@tempc \@tempb \let \@tempb \@tempa \fi \ifx
  \@tempb \@empty \def\@tempb {arXiv}\fi \@ifundefined
  {mn@eprint@\@tempb}{\@tempb:\@tempc}{\expandafter \expandafter \csname
  mn@eprint@\@tempb\endcsname \expandafter{\@tempc}}}

\bibitem[\protect\citeauthoryear{Abazajian}{Abazajian}{2011}]{Abazajian:2010zy}
Abazajian K.~N.,  2011, \mn@doi [JCAP] {10.1088/1475-7516/2011/03/010}, 1103,
  010

\bibitem[\protect\citeauthoryear{Abazajian \& Kaplinghat}{Abazajian \&
  Kaplinghat}{2012}]{Abazajian:2012pn}
Abazajian K.~N.,  Kaplinghat M.,  2012, \mn@doi [Phys.Rev.]
  {10.1103/PhysRevD.86.083511}, D86, 083511

\bibitem[\protect\citeauthoryear{Abazajian, Canac, Horiuchi  \&
  Kaplinghat}{Abazajian et~al.}{2014}]{Abazajian:2014fta}
Abazajian K.~N.,  Canac N.,  Horiuchi S.,   Kaplinghat M.,  2014, \mn@doi
  [Phys. Rev.] {10.1103/PhysRevD.90.023526}, D90, 023526

\bibitem[\protect\citeauthoryear{Acero et~al.}{Acero
  et~al.}{2015}]{Acero:2015hja}
Acero F.,  et~al., 2015

\bibitem[\protect\citeauthoryear{Ajello et~al.}{Ajello
  et~al.}{2016}]{TheFermi-LAT:2015kwa}
Ajello M.,  et~al., 2016, \mn@doi [Astrophys. J.] {10.3847/0004-637X/819/1/44},
  819, 44

\bibitem[\protect\citeauthoryear{Ajello et~al.}{Ajello
  et~al.}{2017}]{Fermi-LAT:2017yoi}
Ajello M.,  et~al., 2017, Submitted to: Astrophys. J.

\bibitem[\protect\citeauthoryear{Alexis, Jean, Martin  \& Ferriere}{Alexis
  et~al.}{2014}]{Alexis:2014rba}
Alexis A.,  Jean P.,  Martin P.,   Ferriere K.,  2014, \mn@doi [Astron.
  Astrophys.] {10.1051/0004-6361/201322393}, 564, A108

\bibitem[\protect\citeauthoryear{Arca-Sedda, Kocsis  \& Brandt}{Arca-Sedda
  et~al.}{2017}]{Arca-Sedda:2017qcq}
Arca-Sedda M.,  Kocsis B.,   Brandt T.,  2017

\bibitem[\protect\citeauthoryear{Archibald et~al.}{Archibald
  et~al.}{2009}]{Archibald:2009zb}
Archibald A.~M.,  et~al., 2009, \mn@doi [Science] {10.1126/science.1172740},
  324, 1411

\bibitem[\protect\citeauthoryear{Archibald et~al.}{Archibald
  et~al.}{2015}]{Archibald:2014nda}
Archibald A.~M.,  et~al., 2015, \mn@doi [Astrophys. J.]
  {10.1088/0004-637X/807/1/62}, 807, 62

\bibitem[\protect\citeauthoryear{Bahramian et~al.}{Bahramian
  et~al.}{2017}]{Bahramian:2017mbs}
Bahramian A.,  et~al., 2017, \mn@doi [Mon. Not. Roy. Astron. Soc.]
  {10.1093/mnras/stx166}, 467, 2199

\bibitem[\protect\citeauthoryear{Bandyopadhyay, Silk, Taylor  \&
  Maccarone}{Bandyopadhyay et~al.}{2009}]{Bandyopadhyay:2008ts}
Bandyopadhyay R.~M.,  Silk J.,  Taylor J.~E.,   Maccarone T.~J.,  2009, \mn@doi
  [Mon. Not. Roy. Astron. Soc.] {10.1111/j.1365-2966.2008.14113.x}, 392, 1115

\bibitem[\protect\citeauthoryear{Bartels, Krishnamurthy  \& Weniger}{Bartels
  et~al.}{2016}]{Bartels:2015aea}
Bartels R.,  Krishnamurthy S.,   Weniger C.,  2016, \mn@doi [Phys. Rev. Lett.]
  {10.1103/PhysRevLett.116.051102}, 116, 051102

\bibitem[\protect\citeauthoryear{Bartels, Hooper, Linden, Mishra-Sharma, Rodd,
  Safdi  \& Slatyer}{Bartels et~al.}{2017b}]{Bartels:2017xba}
Bartels R.,  Hooper D.,  Linden T.,  Mishra-Sharma S.,  Rodd N.~L.,  Safdi
  B.~R.,   Slatyer T.~R.,  2017b

\bibitem[\protect\citeauthoryear{Bartels, Storm, Weniger  \& Calore}{Bartels
  et~al.}{2017a}]{Bartels:2017vsx}
Bartels R.,  Storm E.,  Weniger C.,   Calore F.,  2017a

\bibitem[\protect\citeauthoryear{Bassa et~al.}{Bassa
  et~al.}{2014}]{Bassa:2014hoa}
Bassa C.~G.,  et~al., 2014, \mn@doi [Mon. Not. Roy. Astron. Soc.]
  {10.1093/mnras/stu708}, 441, 1825

\bibitem[\protect\citeauthoryear{Beacom \& Yuksel}{Beacom \&
  Yuksel}{2006}]{Beacom:2005qv}
Beacom J.~F.,  Yuksel H.,  2006, \mn@doi [Phys. Rev. Lett.]
  {10.1103/PhysRevLett.97.071102}, 97, 071102

\bibitem[\protect\citeauthoryear{Belczynski \& Taam}{Belczynski \&
  Taam}{2004}]{Belczynski:2003ac}
Belczynski K.,  Taam R.~E.,  2004, \mn@doi [Astrophys. J.] {10.1086/381491},
  603, 690

\bibitem[\protect\citeauthoryear{Belczynski, Kalogera, Rasio, Taam, Zezas,
  Bulik, Maccarone  \& Ivanova}{Belczynski et~al.}{2008}]{Belczynski:2005mr}
Belczynski K.,  Kalogera V.,  Rasio F.~A.,  Taam R.~E.,  Zezas A.,  Bulik T.,
  Maccarone T.~J.,   Ivanova N.,  2008, \mn@doi [Astrophys. J. Suppl.]
  {10.1086/521026}, 174, 223

\bibitem[\protect\citeauthoryear{{Bhattacharya} \& {van den
  Heuvel}}{{Bhattacharya} \& {van den Heuvel}}{1991}]{1991PhR...203....1B}
{Bhattacharya} D.,  {van den Heuvel} E.~P.~J.,  1991, \mn@doi [\physrep]
  {10.1016/0370-1573(91)90064-S}, \href
  {http://adsabs.harvard.edu/abs/1991PhR...203....1B} {203, 1}

\bibitem[\protect\citeauthoryear{Blandford \& Konigl}{Blandford \&
  Konigl}{1979}]{Blandford:1979za}
Blandford R.~D.,  Konigl A.,  1979, \mn@doi [Astrophys. J.] {10.1086/157262},
  232, 34

\bibitem[\protect\citeauthoryear{Boehm, Hooper, Silk, Casse  \& Paul}{Boehm
  et~al.}{2004}]{Boehm:2003bt}
Boehm C.,  Hooper D.,  Silk J.,  Casse M.,   Paul J.,  2004, \mn@doi [Phys.
  Rev. Lett.] {10.1103/PhysRevLett.92.101301}, 92, 101301

\bibitem[\protect\citeauthoryear{Boehm, Gondolo, Jean, Lacroix, Norman  \&
  Silk}{Boehm et~al.}{2014}]{Boehm:2014bwa}
Boehm C.,  Gondolo P.,  Jean P.,  Lacroix T.,  Norman C.,   Silk J.,  2014

\bibitem[\protect\citeauthoryear{Bogdanov \& Halpern}{Bogdanov \&
  Halpern}{2015}]{Bogdanov:2015xda}
Bogdanov S.,  Halpern J.~P.,  2015, \mn@doi [Astrophys. J.]
  {10.1088/2041-8205/803/2/L27}, 803, L27

\bibitem[\protect\citeauthoryear{Bouchet, Roques  \& Jourdain}{Bouchet
  et~al.}{2010}]{Bouchet:2010dj}
Bouchet L.,  Roques J.-P.,   Jourdain E.,  2010, \mn@doi [Astrophys. J.]
  {10.1088/0004-637X/720/2/1772}, 720, 1772

\bibitem[\protect\citeauthoryear{Brandt \& Kocsis}{Brandt \&
  Kocsis}{2015}]{Brandt:2015ula}
Brandt T.~D.,  Kocsis B.,  2015, \mn@doi [Astrophys. J.]
  {10.1088/0004-637X/812/1/15}, 812, 15

\bibitem[\protect\citeauthoryear{Calore, Cholis, McCabe  \& Weniger}{Calore
  et~al.}{2015a}]{Calore:2014nla}
Calore F.,  Cholis I.,  McCabe C.,   Weniger C.,  2015a, \mn@doi [Phys. Rev.]
  {10.1103/PhysRevD.91.063003}, D91, 063003

\bibitem[\protect\citeauthoryear{Calore, Cholis  \& Weniger}{Calore
  et~al.}{2015b}]{Calore:2014xka}
Calore F.,  Cholis I.,   Weniger C.,  2015b, \mn@doi [JCAP]
  {10.1088/1475-7516/2015/03/038}, 1503, 038

\bibitem[\protect\citeauthoryear{Calore, Di~Mauro, Donato, Hessels  \&
  Weniger}{Calore et~al.}{2016}]{Calore:2015bsx}
Calore F.,  Di~Mauro M.,  Donato F.,  Hessels J. W.~T.,   Weniger C.,  2016,
  \mn@doi [Astrophys. J.] {10.3847/0004-637X/827/2/143}, 827, 143

\bibitem[\protect\citeauthoryear{Cao, Mao, Nataf, Rattenbury  \& Gould}{Cao
  et~al.}{2013}]{Cao:2013dwa}
Cao L.,  Mao S.,  Nataf D.,  Rattenbury N.~J.,   Gould A.,  2013, \mn@doi [Mon.
  Not. Roy. Astron. Soc.] {10.1093/mnras/stt1045}, 434, 595

\bibitem[\protect\citeauthoryear{Caron, G\'{o}mez-Vargas, Hendriks  \& Ruiz~de
  Austri}{Caron et~al.}{2017}]{Caron:2017udl}
Caron S.,  G\'{o}mez-Vargas G.~A.,  Hendriks L.,   Ruiz~de Austri R.,  2017

\bibitem[\protect\citeauthoryear{Chattopadhyay}{Chattopadhyay}{2005}]{Chattopadhyay:2004gf}
Chattopadhyay I.,  2005, \mn@doi [Mon. Not. Roy. Astron. Soc.]
  {10.1111/j.1365-2966.2004.08429.x}, 356, 145

\bibitem[\protect\citeauthoryear{Cholis, Hooper  \& Linden}{Cholis
  et~al.}{2015}]{Cholis:2014lta}
Cholis I.,  Hooper D.,   Linden T.,  2015, \mn@doi [JCAP]
  {10.1088/1475-7516/2015/06/043}, 1506, 043

\bibitem[\protect\citeauthoryear{{Chomiuk}, {Strader}, {Maccarone},
  {Miller-Jones}, {Heinke}, {Noyola}, {Seth}  \& {Ransom}}{{Chomiuk}
  et~al.}{2013}]{2013ApJ...777...69C}
{Chomiuk} L.,  {Strader} J.,  {Maccarone} T.~J.,  {Miller-Jones} J.~C.~A.,
  {Heinke} C.,  {Noyola} E.,  {Seth} A.~C.,   {Ransom} S.,  2013, \mn@doi
  [\apj] {10.1088/0004-637X/777/1/69}, \href
  {http://adsabs.harvard.edu/abs/2013ApJ...777...69C} {777, 69}

\bibitem[\protect\citeauthoryear{Churazov, Sunyaev, Sazonov, Revnivtsev  \&
  Varshalovich}{Churazov et~al.}{2005}]{Churazov:2004as}
Churazov E.,  Sunyaev R.,  Sazonov S.,  Revnivtsev M.,   Varshalovich D.,
  2005, \mn@doi [Mon. Not. Roy. Astron. Soc.]
  {10.1111/j.1365-2966.2005.08757.x}, 357, 1377

\bibitem[\protect\citeauthoryear{{Clayton}}{{Clayton}}{1973}]{1973NPhS..244R.137C}
{Clayton} D.~D.,  1973, \mn@doi [Nature Physical Science]
  {10.1038/physci244137b0}, \href
  {http://adsabs.harvard.edu/abs/1973NPhS..244R.137C} {244, 137}

\bibitem[\protect\citeauthoryear{Corbel, Nowak, Fender, Tzioumis  \&
  Markoff}{Corbel et~al.}{2003}]{Corbel:2003wt}
Corbel S.,  Nowak M.~A.,  Fender R.~P.,  Tzioumis A.~K.,   Markoff S.,  2003,
  \mn@doi [Astron. Astrophys.] {10.1051/0004-6361:20030090}, 400, 1007

\bibitem[\protect\citeauthoryear{Crocker et~al.}{Crocker
  et~al.}{2016}]{Crocker:2016zzt}
Crocker R.~M.,  et~al., 2016, ] {10.1038/s41550-017-0135}

\bibitem[\protect\citeauthoryear{Daylan, Finkbeiner, Hooper, Linden, Portillo,
  Rodd  \& Slatyer}{Daylan et~al.}{2016}]{Daylan:2014rsa}
Daylan T.,  Finkbeiner D.~P.,  Hooper D.,  Linden T.,  Portillo S. K.~N.,  Rodd
  N.~L.,   Slatyer T.~R.,  2016, \mn@doi [Phys. Dark Univ.]
  {10.1016/j.dark.2015.12.005}, 12, 1

\bibitem[\protect\citeauthoryear{De~Angelis et~al.}{De~Angelis
  et~al.}{2016}]{DeAngelis:2016slk}
De~Angelis A.,  et~al., 2016, ] {10.1007/s10686-017-9533-6}

\bibitem[\protect\citeauthoryear{De~Cesare}{De~Cesare}{2011}]{DeCesare:2011gc}
De~Cesare G.,  2011, \mn@doi [Astron. Astrophys.]
  {10.1051/0004-6361/201116516}, 531, A56

\bibitem[\protect\citeauthoryear{Deller et~al.,}{Deller
  et~al.}{2015}]{Deller:2014yca}
Deller A.~T.,  et~al., 2015, \mn@doi [Astrophys. J.]
  {10.1088/0004-637X/809/1/13}, 809, 13

\bibitem[\protect\citeauthoryear{Done, Gierlinski  \& Kubota}{Done
  et~al.}{2007}]{Done:2007nc}
Done C.,  Gierlinski M.,   Kubota A.,  2007, \mn@doi [Astron. Astrophys. Rev.]
  {10.1007/s00159-007-0006-1}, 15, 1

\bibitem[\protect\citeauthoryear{Falcke \& Biermann}{Falcke \&
  Biermann}{1995}]{Falcke:1994eb}
Falcke H.,  Biermann P.~L.,  1995, Astron. Astrophys., 293, 665

\bibitem[\protect\citeauthoryear{Falcke, Koerding  \& Markoff}{Falcke
  et~al.}{2004}]{Falcke:2003ia}
Falcke H.,  Koerding E.,   Markoff S.,  2004, \mn@doi [Astron. Astrophys.]
  {10.1051/0004-6361:20031683}, 414, 895

\bibitem[\protect\citeauthoryear{Fender, Gallo  \& Jonker}{Fender
  et~al.}{2003}]{Fender:2003ae}
Fender R.~P.,  Gallo E.,   Jonker P.~G.,  2003, \mn@doi [Mon. Not. Roy. Astron.
  Soc.] {10.1046/j.1365-8711.2003.06950.x}, 343, L99

\bibitem[\protect\citeauthoryear{Fender, Belloni  \& Gallo}{Fender
  et~al.}{2004}]{Fender:2004gg}
Fender R.~P.,  Belloni T.~M.,   Gallo E.,  2004, \mn@doi [Mon. Not. Roy.
  Astron. Soc.] {10.1111/j.1365-2966.2004.08384.x}, 355, 1105

\bibitem[\protect\citeauthoryear{Fender, Maccarone  \& van Kesteren}{Fender
  et~al.}{2005}]{Fender:2005nh}
Fender R.,  Maccarone T.,   van Kesteren Z.,  2005, \mn@doi [Mon. Not. Roy.
  Astron. Soc.] {10.1111/j.1365-2966.2005.09098.x}, 360, 1085

\bibitem[\protect\citeauthoryear{Finkbeiner \& Weiner}{Finkbeiner \&
  Weiner}{2007}]{Finkbeiner:2007kk}
Finkbeiner D.~P.,  Weiner N.,  2007, \mn@doi [Phys. Rev.]
  {10.1103/PhysRevD.76.083519}, D76, 083519

\bibitem[\protect\citeauthoryear{Fragione, Antonini  \& Gnedin}{Fragione
  et~al.}{2017}]{Fragione:2017rsp}
Fragione G.,  Antonini F.,   Gnedin O.~Y.,  2017, ] {10.1093/mnras/sty183}

\bibitem[\protect\citeauthoryear{{Frank}, {King}  \& {Raine}}{{Frank}
  et~al.}{1992}]{1992apa..book.....F}
{Frank} J.,  {King} A.,   {Raine} D.,  1992, {Accretion power in astrophysics.}

\bibitem[\protect\citeauthoryear{Frey \& Reid}{Frey \&
  Reid}{2013}]{Frey:2013wh}
Frey A.~R.,  Reid N.~B.,  2013, \mn@doi [Phys. Rev.]
  {10.1103/PhysRevD.87.103508}, D87, 103508

\bibitem[\protect\citeauthoryear{Gallo, Fender  \& Pooley}{Gallo
  et~al.}{2003}]{Gallo:2003tv}
Gallo E.,  Fender R.~P.,   Pooley G.~G.,  2003, \mn@doi [Mon. Not. Roy. Astron.
  Soc.] {10.1046/j.1365-8711.2003.06791.x}, 344, 60

\bibitem[\protect\citeauthoryear{Gallo, Fender, Kaiser, Russell, Morganti,
  Oosterloo  \& Heinz}{Gallo et~al.}{2005}]{Gallo:2005tf}
Gallo E.,  Fender R.,  Kaiser C.,  Russell D.,  Morganti R.,  Oosterloo T.,
  Heinz S.,  2005, \mn@doi [Nature] {10.1038/nature03879}, 436, 819

\bibitem[\protect\citeauthoryear{Gallo, Fender, Miller-Jones, Merloni, Jonker,
  Heinz, Maccarone  \& van~der Klis}{Gallo et~al.}{2006}]{Gallo:2006dz}
Gallo E.,  Fender R.,  Miller-Jones J.,  Merloni A.,  Jonker P.,  Heinz S.,
  Maccarone T.,   van~der Klis M.,  2006, \mn@doi [Mon. Not. Roy. Astron. Soc.]
  {10.1111/j.1365-2966.2006.10560.x}, 370, 1351

\bibitem[\protect\citeauthoryear{Gallo et~al.,}{Gallo
  et~al.}{2014}]{Gallo:2014lba}
Gallo E.,  et~al., 2014, \mn@doi [Mon. Not. Roy. Astron. Soc.]
  {10.1093/mnras/stu1599}, 445, 290

\bibitem[\protect\citeauthoryear{Gilfanov}{Gilfanov}{2004}]{Gilfanov:2003th}
Gilfanov M.,  2004, \mn@doi [Mon. Not. Roy. Astron. Soc.]
  {10.1111/j.1365-2966.2004.07473.x}, 349, 146

\bibitem[\protect\citeauthoryear{Goldoni, Goldwurm, Laurent  \& Lebrun}{Goldoni
  et~al.}{1997}]{Goldoni:1997ti}
Goldoni P.,  Goldwurm A.,  Laurent P.,   Lebrun F.,  1997, \mn@doi [AIP Conf.
  Proc.] {10.1063/1.54071}, 410, 1549

\bibitem[\protect\citeauthoryear{Goldwurm et~al.,}{Goldwurm
  et~al.}{2003}]{Goldwurm:2003vx}
Goldwurm A.,  et~al., 2003, \mn@doi [Astron. Astrophys.]
  {10.1051/0004-6361:20031395}, 411, L223

\bibitem[\protect\citeauthoryear{Gonzalez-Nuevo, Argueso, Lopez-Caniego,
  Toffolatti, Sanz, Vielva  \& Herranz}{Gonzalez-Nuevo
  et~al.}{2006}]{GonzalezNuevo:2006pa}
Gonzalez-Nuevo J.,  Argueso F.,  Lopez-Caniego M.,  Toffolatti L.,  Sanz J.~L.,
   Vielva P.,   Herranz D.,  2006, \mn@doi [Mon. Not. Roy. Astron. Soc.]
  {10.1111/j.1365-2966.2006.10442.x}, 369, 1603

\bibitem[\protect\citeauthoryear{Goodenough \& Hooper}{Goodenough \&
  Hooper}{2009}]{Goodenough:2009gk}
Goodenough L.,  Hooper D.,  2009

\bibitem[\protect\citeauthoryear{Gordon \& Macias}{Gordon \&
  Macias}{2013}]{Gordon:2013vta}
Gordon C.,  Macias O.,  2013, \mn@doi [Phys. Rev.] {10.1103/PhysRevD.88.083521,
  10.1103/PhysRevD.89.049901}, D88, 083521

\bibitem[\protect\citeauthoryear{Guessoum, Jean  \& Gillard}{Guessoum
  et~al.}{2005}]{Guessoum2005}
Guessoum N.,  Jean P.,   Gillard W.,  2005, \mn@doi [Astron. Astrophys.]
  {10.1051/0004-6361:20042454}

\bibitem[\protect\citeauthoryear{Guessoum, Jean  \& Prantzos}{Guessoum
  et~al.}{2006}]{Guessoum:2006fs}
Guessoum N.,  Jean P.,   Prantzos N.,  2006, \mn@doi [Astron. Astrophys.]
  {10.1051/0004-6361:20065240}

\bibitem[\protect\citeauthoryear{Haggard, Heinke, Hooper  \& Linden}{Haggard
  et~al.}{2017}]{Haggard:2017lyq}
Haggard D.,  Heinke C.,  Hooper D.,   Linden T.,  2017, \mn@doi [JCAP]
  {10.1088/1475-7516/2017/05/056}, 1705, 056

\bibitem[\protect\citeauthoryear{{Hanawa}}{{Hanawa}}{1989}]{1989ApJ...341..948H}
{Hanawa} T.,  1989, \mn@doi [\apj] {10.1086/167553}, \href
  {http://adsabs.harvard.edu/abs/1989ApJ...341..948H} {341, 948}

\bibitem[\protect\citeauthoryear{Heinz}{Heinz}{2005}]{Heinz:2005jc}
Heinz S.,  2005, \mn@doi [Astrophys. J.] {10.1086/497954}, 636, 316

\bibitem[\protect\citeauthoryear{Heinz \& Grimm}{Heinz \&
  Grimm}{2005}]{Heinz:2005hw}
Heinz S.,  Grimm H.~J.,  2005, \mn@doi [Astrophys. J.] {10.1086/452624}, 633,
  384

\bibitem[\protect\citeauthoryear{Heinz \& Sunyaev}{Heinz \&
  Sunyaev}{2002}]{Heinz:2002qj}
Heinz S.,  Sunyaev R.~A.,  2002, \mn@doi [Astron. Astrophys.]
  {10.1051/0004-6361:20020615}, 390, 751

\bibitem[\protect\citeauthoryear{Heinz \& Sunyaev}{Heinz \&
  Sunyaev}{2003}]{Heinz:2003xt}
Heinz S.,  Sunyaev R.~A.,  2003, \mn@doi [Mon. Not. Roy. Astron. Soc.]
  {10.1046/j.1365-8711.2003.06918.x}, 343, L59

\bibitem[\protect\citeauthoryear{Hong et~al.}{Hong et~al.}{2016}]{Hong:2016qjq}
Hong J.,  et~al., 2016, \mn@doi [Astrophys. J.] {10.3847/0004-637X/825/2/132},
  825, 132

\bibitem[\protect\citeauthoryear{Hooper \& Goodenough}{Hooper \&
  Goodenough}{2011}]{Hooper:2010mq}
Hooper D.,  Goodenough L.,  2011, \mn@doi [Phys. Lett.]
  {10.1016/j.physletb.2011.02.029}, B697, 412

\bibitem[\protect\citeauthoryear{Hooper \& Linden}{Hooper \&
  Linden}{2011}]{Hooper:2011ti}
Hooper D.,  Linden T.,  2011, \mn@doi [Phys. Rev.]
  {10.1103/PhysRevD.84.123005}, D84, 123005

\bibitem[\protect\citeauthoryear{Hooper, Cholis, Linden, Siegal-Gaskins  \&
  Slatyer}{Hooper et~al.}{2013}]{Hooper:2013nhl}
Hooper D.,  Cholis I.,  Linden T.,  Siegal-Gaskins J.,   Slatyer T.,  2013,
  \mn@doi [Phys. Rev.] {10.1103/PhysRevD.88.083009}, D88, 083009

\bibitem[\protect\citeauthoryear{Ivanova, Heinke, Rasio, Belczynski  \&
  Fregeau}{Ivanova et~al.}{2008}]{Ivanova:2007bu}
Ivanova N.,  Heinke C.,  Rasio F.~A.,  Belczynski K.,   Fregeau J.,  2008,
  \mn@doi [Mon. Not. Roy. Astron. Soc.] {10.1111/j.1365-2966.2008.13064.x},
  386, 553

\bibitem[\protect\citeauthoryear{Jaodand, Archibald, Hessels, Bogdanov,
  D'Angelo, Patruno, Bassa  \& Deller}{Jaodand et~al.}{2016}]{Jaodand:2016hry}
Jaodand A.,  Archibald A.~M.,  Hessels J. W.~T.,  Bogdanov S.,  D'Angelo C.~R.,
   Patruno A.,  Bassa C.,   Deller A.~T.,  2016, \mn@doi [Astrophys. J.]
  {10.3847/0004-637X/830/2/122}, 830, 122

\bibitem[\protect\citeauthoryear{Jean, Knodlseder, Gillard, Guessoum, Ferriere,
  Marcowith, Lonjou  \& Roques}{Jean et~al.}{2006}]{Jean:2005af}
Jean P.,  Knodlseder J.,  Gillard W.,  Guessoum N.,  Ferriere K.,  Marcowith
  A.,  Lonjou V.,   Roques J.~P.,  2006, \mn@doi [Astron. Astrophys.]
  {10.1051/0004-6361:20053765}, 445, 579

\bibitem[\protect\citeauthoryear{Jonker et~al.}{Jonker
  et~al.}{2011}]{Jonker:2011gf}
Jonker P.~G.,  et~al., 2011, \mn@doi [Astrophys. J. Suppl.]
  {10.1088/0067-0049/194/2/18}, 194, 18

\bibitem[\protect\citeauthoryear{Jourdain, Roques  \& Malzac}{Jourdain
  et~al.}{2012}]{Jourdain:2011we}
Jourdain E.,  Roques J.-P.,   Malzac J.,  2012, \mn@doi [Astrophys. J.]
  {10.1088/0004-637X/744/1/64}, 744, 64

\bibitem[\protect\citeauthoryear{Knodlseder et~al.}{Knodlseder
  et~al.}{2005}]{Knodlseder:2005yq}
Knodlseder J.,  et~al., 2005, \mn@doi [Astron. Astrophys.]
  {10.1051/0004-6361:20042063}, 441, 513

\bibitem[\protect\citeauthoryear{Kording, Fender  \& Migliari}{Kording
  et~al.}{2006}]{Kording:2006sa}
Kording E.,  Fender R.,   Migliari S.,  2006, \mn@doi [Mon. Not. Roy. Astron.
  Soc.] {10.1111/j.1365-2966.2006.10383.x}, 369, 1451

\bibitem[\protect\citeauthoryear{Launhardt, Zylka  \& Mezger}{Launhardt
  et~al.}{2002}]{Launhardt:2002tx}
Launhardt R.,  Zylka R.,   Mezger P.~G.,  2002, \mn@doi [Astron. Astrophys.]
  {10.1051/0004-6361:20020017}, 384, 112

\bibitem[\protect\citeauthoryear{Lebrun et~al.}{Lebrun
  et~al.}{2003}]{Lebrun:2003aa}
Lebrun F.,  et~al., 2003, \mn@doi [Astron. Astrophys.]
  {10.1051/0004-6361:20031367}, 411, L141

\bibitem[\protect\citeauthoryear{Lee, Lisanti, Safdi, Slatyer  \& Xue}{Lee
  et~al.}{2016}]{Lee:2015fea}
Lee S.~K.,  Lisanti M.,  Safdi B.~R.,  Slatyer T.~R.,   Xue W.,  2016, \mn@doi
  [Phys. Rev. Lett.] {10.1103/PhysRevLett.116.051103}, 116, 051103

\bibitem[\protect\citeauthoryear{{Leventhal}, {MacCallum}  \&
  {Stang}}{{Leventhal} et~al.}{1978}]{1978ApJ...225L..11L}
{Leventhal} M.,  {MacCallum} C.~J.,   {Stang} P.~D.,  1978, \mn@doi [\apjl]
  {10.1086/182782}, \href {http://adsabs.harvard.edu/abs/1978ApJ...225L..11L}
  {225, L11}

\bibitem[\protect\citeauthoryear{Licquia \& Newman}{Licquia \&
  Newman}{2015}]{Licquia:2014rsa}
Licquia T.~C.,  Newman J.~A.,  2015, \mn@doi [Astrophys. J.]
  {10.1088/0004-637X/806/1/96}, 806, 96

\bibitem[\protect\citeauthoryear{Linares et~al.,}{Linares
  et~al.}{2014}]{Linares:2013ewa}
Linares M.,  et~al., 2014, \mn@doi [Mon. Not. Roy. Astron. Soc.]
  {10.1093/mnras/stt2167}, 438, 251

\bibitem[\protect\citeauthoryear{{Liu}, {van Paradijs}  \& {van den
  Heuvel}}{{Liu} et~al.}{2007}]{2007A&A...469..807L}
{Liu} Q.~Z.,  {van Paradijs} J.,   {van den Heuvel} E.~P.~J.,  2007, \mn@doi
  [\aap] {10.1051/0004-6361:20077303}, \href
  {http://adsabs.harvard.edu/abs/2007A%26A...469..807L} {469, 807}

\bibitem[\protect\citeauthoryear{{Maccarone} \& {Peacock}}{{Maccarone} \&
  {Peacock}}{2011}]{2011MNRAS.415.1875M}
{Maccarone} T.~J.,  {Peacock} M.~B.,  2011, \mn@doi [\mnras]
  {10.1111/j.1365-2966.2011.18831.x}, \href
  {http://adsabs.harvard.edu/abs/2011MNRAS.415.1875M} {415, 1875}

\bibitem[\protect\citeauthoryear{{Maccarone}, {Kundu}, {Zepf}  \&
  {Rhode}}{{Maccarone} et~al.}{2007}]{2007Natur.445..183M}
{Maccarone} T.~J.,  {Kundu} A.,  {Zepf} S.~E.,   {Rhode} K.~L.,  2007, \mn@doi
  [\nat] {10.1038/nature05434}, \href
  {http://adsabs.harvard.edu/abs/2007Natur.445..183M} {445, 183}

\bibitem[\protect\citeauthoryear{Macias \& Gordon}{Macias \&
  Gordon}{2014}]{Macias:2013vya}
Macias O.,  Gordon C.,  2014, \mn@doi [Phys.Rev.] {10.1103/PhysRevD.89.063515},
  D89, 063515

\bibitem[\protect\citeauthoryear{Macias, Gordon, Crocker, Coleman, Paterson,
  Horiuchi  \& Pohl}{Macias et~al.}{2018}]{Macias:2016nev}
Macias O.,  Gordon C.,  Crocker R.~M.,  Coleman B.,  Paterson D.,  Horiuchi S.,
    Pohl M.,  2018, \mn@doi [Nature Astronomy] {10.1038/s41550-018-0414-3}

\bibitem[\protect\citeauthoryear{Markoff, Falcke  \& Fender}{Markoff
  et~al.}{2001}]{Markoff:2000ir}
Markoff S.,  Falcke H.,   Fender R.~P.,  2001, \mn@doi [Astron. Astrophys.]
  {10.1051/0004-6361:20010420}, 372, L25

\bibitem[\protect\citeauthoryear{Markoff, Nowak, Corbel, Fender  \&
  Falcke}{Markoff et~al.}{2003}]{Markoff:2002xs}
Markoff S.,  Nowak M.,  Corbel S.,  Fender R.~P.,   Falcke H.,  2003, \mn@doi
  [Astron. Astrophys.] {10.1051/0004-6361:20021497}, 397, 645

\bibitem[\protect\citeauthoryear{Merloni, Heinz  \& Di~Matteo}{Merloni
  et~al.}{2003}]{Merloni:2003aq}
Merloni A.,  Heinz S.,   Di~Matteo T.,  2003, \mn@doi [Mon. Not. Roy. Astron.
  Soc.] {10.1046/j.1365-2966.2003.07017.x}, 345, 1057

\bibitem[\protect\citeauthoryear{Migliari \& Fender}{Migliari \&
  Fender}{2006}]{Migliari:2005hv}
Migliari S.,  Fender R.~P.,  2006, \mn@doi [Mon. Not. Roy. Astron. Soc.]
  {10.1111/j.1365-2966.2005.09777.x}, 366, 79

\bibitem[\protect\citeauthoryear{Migliari, Fender  \& Mendez}{Migliari
  et~al.}{2002}]{Migliari:2002tc}
Migliari S.,  Fender R.~P.,   Mendez M.,  2002, \mn@doi [Science]
  {10.1126/science.1073660}, 297, 1673

\bibitem[\protect\citeauthoryear{{Miller-Jones} et~al.,}{{Miller-Jones}
  et~al.}{2015a}]{2015MNRAS.453.3918M}
{Miller-Jones} J.~C.~A.,  et~al., 2015a, \mn@doi [\mnras]
  {10.1093/mnras/stv1869}, \href
  {http://adsabs.harvard.edu/abs/2015MNRAS.453.3918M} {453, 3918}

\bibitem[\protect\citeauthoryear{Miller-Jones et~al.}{Miller-Jones
  et~al.}{2015b}]{Miller-Jones:2015zba}
Miller-Jones J. C.~A.,  et~al., 2015b, \mn@doi [Mon. Not. Roy. Astron. Soc.]
  {10.1093/mnras/stv1869}, 453, 3918

\bibitem[\protect\citeauthoryear{Mirabel \& Rodriguez}{Mirabel \&
  Rodriguez}{1999}]{Mirabel:1999fy}
Mirabel I.~F.,  Rodriguez L.~F.,  1999, \mn@doi [Ann. Rev. Astron. Astrophys.]
  {10.1146/annurev.astro.37.1.409}, 37, 409

\bibitem[\protect\citeauthoryear{Narayan \& Yi}{Narayan \&
  Yi}{1994}]{Narayan:1994xi}
Narayan R.,  Yi I.-s.,  1994, \mn@doi [Astrophys. J.] {10.1086/187381}, 428,
  L13

\bibitem[\protect\citeauthoryear{Navarro, Frenk  \& White}{Navarro
  et~al.}{1997}]{Navarro:1996gj}
Navarro J.~F.,  Frenk C.~S.,   White S. D.~M.,  1997, \mn@doi [Astrophys. J.]
  {10.1086/304888}, 490, 493

\bibitem[\protect\citeauthoryear{{Nolan} et~al.,}{{Nolan}
  et~al.}{2012}]{Fermi-LAT:2011yjw}
{Nolan} P.~L.,  et~al., 2012, \mn@doi [Astrophys. J. Suppl.]
  {10.1088/0067-0049/199/2/31}, 199, 31

\bibitem[\protect\citeauthoryear{Papitto et~al.}{Papitto
  et~al.}{2013}]{Papitto:2013hza}
Papitto A.,  et~al., 2013, \mn@doi [Nature] {10.1038/nature12470}, 501, 517

\bibitem[\protect\citeauthoryear{Papitto, de Martino, Belloni, Burgay,
  Pellizzoni, Possenti  \& Torres}{Papitto et~al.}{2015}]{Papitto:2014rua}
Papitto A.,  de Martino D.,  Belloni T.~M.,  Burgay M.,  Pellizzoni A.,
  Possenti A.,   Torres D.~F.,  2015, \mn@doi [Mon. Not. Roy. Astron. Soc.]
  {10.1093/mnrasl/slv013}, 449, L26

\bibitem[\protect\citeauthoryear{Petrovi\'{c}, Serpico  \&
  Zaharijas}{Petrovi\'{c} et~al.}{2015}]{Petrovic:2014xra}
Petrovi\'{c} J.,  Serpico P.~D.,   Zaharijas G.,  2015, \mn@doi [JCAP]
  {10.1088/1475-7516/2015/02/023}, 1502, 023

\bibitem[\protect\citeauthoryear{Ploeg, Gordon, Crocker  \& Macias}{Ploeg
  et~al.}{2017}]{Ploeg:2017vai}
Ploeg H.,  Gordon C.,  Crocker R.,   Macias O.,  2017, \mn@doi [JCAP]
  {10.1088/1475-7516/2017/08/015}, 1708, 015

\bibitem[\protect\citeauthoryear{{Prantzos}}{{Prantzos}}{2004}]{2004ESASP.552...15P}
{Prantzos} N.,  2004, in {Schoenfelder} V.,  {Lichti} G.,   {Winkler} C.,  eds,
   ESA Special Publication Vol. 552, 5th INTEGRAL Workshop on the INTEGRAL
  Universe. p.~15 (\mn@eprint {} {astro-ph/0404501})

\bibitem[\protect\citeauthoryear{Prantzos et~al.}{Prantzos
  et~al.}{2011}]{Prantzos:2010wi}
Prantzos N.,  et~al., 2011, \mn@doi [Rev. Mod. Phys.]
  {10.1103/RevModPhys.83.1001}, 83, 1001

\bibitem[\protect\citeauthoryear{Remillard \& McClintock}{Remillard \&
  McClintock}{2006}]{Remillard:2006fc}
Remillard R.~A.,  McClintock J.~E.,  2006, \mn@doi [Ann. Rev. Astron.
  Astrophys.] {10.1146/annurev.astro.44.051905.092532}, 44, 49

\bibitem[\protect\citeauthoryear{Romanova, Ustyugova, Koldoba  \&
  Lovelace}{Romanova et~al.}{2004}]{Romanova:2005di}
Romanova M.~M.,  Ustyugova G.~V.,  Koldoba A.~V.,   Lovelace R. V.~E.,  2004,
  \mn@doi [Astrophys. J.] {10.1086/426586}, 616, L151

\bibitem[\protect\citeauthoryear{Romero, Christiansen  \& Orellana}{Romero
  et~al.}{2005}]{Romero:2005fr}
Romero G.~E.,  Christiansen H.~R.,   Orellana M.,  2005, \mn@doi [Astrophys.
  J.] {10.1086/444446}, 632, 1093

\bibitem[\protect\citeauthoryear{Roques, Jourdain, Bazzano, Fiocchi, Natalucci
  \& Ubertini}{Roques et~al.}{2015}]{Roques:2015bma}
Roques J.-P.,  Jourdain E.,  Bazzano A.,  Fiocchi M.,  Natalucci L.,   Ubertini
  P.,  2015, \mn@doi [Astrophys. J.] {10.1088/2041-8205/813/1/L22}, 813, L22

\bibitem[\protect\citeauthoryear{Roy et~al.}{Roy et~al.}{2015}]{Roy:2014cwa}
Roy J.,  et~al., 2015, \mn@doi [Astrophys. J.] {10.1088/2041-8205/800/1/L12},
  800, L12

\bibitem[\protect\citeauthoryear{Shakura \& Sunyaev}{Shakura \&
  Sunyaev}{1976}]{Shakura:1976xk}
Shakura N.~I.,  Sunyaev R.~A.,  1976, Mon. Not. Roy. Astron. Soc., 175, 613

\bibitem[\protect\citeauthoryear{Siegert et~al.,}{Siegert
  et~al.}{2016a}]{Siegert:2016ymf}
Siegert T.,  et~al., 2016a, \mn@doi [Nature] {10.1038/nature16978}, 341

\bibitem[\protect\citeauthoryear{Siegert, Diehl, Khachatryan, Krause,
  Guglielmetti, Greiner, Strong  \& Zhang}{Siegert
  et~al.}{2016b}]{Siegert:2015knp}
Siegert T.,  Diehl R.,  Khachatryan G.,  Krause M. G.~H.,  Guglielmetti F.,
  Greiner J.,  Strong A.~W.,   Zhang X.,  2016b, \mn@doi [Astron. Astrophys.]
  {10.1051/0004-6361/201527510}, 586, A84

\bibitem[\protect\citeauthoryear{Siegert, Diehl, Vincent, Guglielmetti, Krause
  \& Boehm}{Siegert et~al.}{2016c}]{Siegert:2016ijv}
Siegert T.,  Diehl R.,  Vincent A.~C.,  Guglielmetti F.,  Krause M. G.~H.,
  Boehm C.,  2016c, \mn@doi [Astron. Astrophys.] {10.1051/0004-6361/201629136},
  595, A25

\bibitem[\protect\citeauthoryear{Sizun, Casse  \& Schanne}{Sizun
  et~al.}{2006}]{Sizun:2006uh}
Sizun P.,  Casse M.,   Schanne S.,  2006, \mn@doi [Phys. Rev.]
  {10.1103/PhysRevD.74.063514}, D74, 063514

\bibitem[\protect\citeauthoryear{{Skinner}, Diehl, Zhang, Bouchet  \&
  Jean}{{Skinner} et~al.}{2014}]{PoS(INTEGRAL2014)054}
{Skinner} G.,  Diehl R.,  Zhang X.~L.,  Bouchet L.,   Jean P.,  2014, in
  Proceedings of the 10th INTEGRAL Workshop: A Synergistic View of the
  High-Energy Sky. 15-19 September 2014. Annapolis, MD, USA. Published online
  at \url{https://pos.sissa.it/228/}..

\bibitem[\protect\citeauthoryear{Storm, Weniger  \& Calore}{Storm
  et~al.}{2017}]{Storm:2017arh}
Storm E.,  Weniger C.,   Calore F.,  2017

\bibitem[\protect\citeauthoryear{{Strader}, {Chomiuk}, {Maccarone},
  {Miller-Jones}  \& {Seth}}{{Strader} et~al.}{2012}]{2012Natur.490...71S}
{Strader} J.,  {Chomiuk} L.,  {Maccarone} T.~J.,  {Miller-Jones} J.~C.~A.,
  {Seth} A.~C.,  2012, \mn@doi [\nat] {10.1038/nature11490}, \href
  {http://adsabs.harvard.edu/abs/2012Natur.490...71S} {490, 71}

\bibitem[\protect\citeauthoryear{Strong, Diehl, Halloin, Schoenfelder, Bouchet,
  Mandrou, Lebrun  \& Terrier}{Strong et~al.}{2005}]{Strong:2005zx}
Strong A.~W.,  Diehl R.,  Halloin H.,  Schoenfelder V.,  Bouchet L.,  Mandrou
  P.,  Lebrun F.,   Terrier R.,  2005, \mn@doi [Astron. Astrophys.]
  {10.1051/0004-6361:20053798}, 444, 495

\bibitem[\protect\citeauthoryear{Tavani et~al.}{Tavani
  et~al.}{2017}]{DeAngelis:2017gra}
Tavani M.,  et~al., 2017

\bibitem[\protect\citeauthoryear{Tendulkar et~al.}{Tendulkar
  et~al.}{2014}]{Tendulkar:2014wga}
Tendulkar S.~P.,  et~al., 2014, \mn@doi [Astrophys. J.]
  {10.1088/0004-637X/791/2/77}, 791, 77

\bibitem[\protect\citeauthoryear{Tetarenko, Sivakoff, Heinke  \&
  Gladstone}{Tetarenko et~al.}{2016a}]{Tetarenko:2015vrn}
Tetarenko B.~E.,  Sivakoff G.~R.,  Heinke C.~O.,   Gladstone J.~C.,  2016a,
  \mn@doi [Astrophys. J. Suppl.] {10.3847/0067-0049/222/2/15}, 222, 15

\bibitem[\protect\citeauthoryear{Tetarenko et~al.}{Tetarenko
  et~al.}{2016b}]{Tetarenko:2016uln}
Tetarenko B.~E.,  et~al., 2016b, \mn@doi [Astrophys. J.]
  {10.3847/0004-637X/825/1/10}, 825, 10

\bibitem[\protect\citeauthoryear{Totani}{Totani}{2006}]{Totani:2006zx}
Totani T.,  2006, \mn@doi [Publ. Astron. Soc. Jap.] {10.1093/pasj/58.6.965},
  58, 965

\bibitem[\protect\citeauthoryear{Ubertini et~al.}{Ubertini
  et~al.}{2003}]{Ubertini:2003ih}
Ubertini P.,  et~al., 2003, \mn@doi [Astron. Astrophys.]
  {10.1051/0004-6361:20031224}, 411, L131

\bibitem[\protect\citeauthoryear{Vincent, Martin  \& Cline}{Vincent
  et~al.}{2012}]{Vincent:2012an}
Vincent A.~C.,  Martin P.,   Cline J.~M.,  2012, \mn@doi [JCAP]
  {10.1088/1475-7516/2012/04/022}, 1204, 022

\bibitem[\protect\citeauthoryear{Vitale \& Morselli}{Vitale \&
  Morselli}{2009}]{Vitale:2009hr}
Vitale V.,  Morselli A.,  2009, in {Fermi gamma-ray space telescope.
  Proceedings, 2nd Fermi Symposium, Washington, USA, November 2-5, 2009}.
  (\mn@eprint {arXiv} {0912.3828}), \url
  {https://inspirehep.net/record/840760/files/arXiv:0912.3828.pdf}

\bibitem[\protect\citeauthoryear{Wang, Pun  \& Cheng}{Wang
  et~al.}{2006}]{Wang:2005cqa}
Wang W.,  Pun C. S.~J.,   Cheng K.~S.,  2006, \mn@doi [Astron. Astrophys.]
  {10.1051/0004-6361:20053559}, 446, 943

\bibitem[\protect\citeauthoryear{Weidenspointner et~al.}{Weidenspointner
  et~al.}{2008}]{Weidenspointner:2008zz}
Weidenspointner G.,  et~al., 2008, \mn@doi [Nature] {10.1038/nature06490}, 451,
  159

\bibitem[\protect\citeauthoryear{Wilkinson, Vincent, Boehm  \&
  McCabe}{Wilkinson et~al.}{2016}]{Wilkinson:2016gsy}
Wilkinson R.~J.,  Vincent A.~C.,  Boehm C.,   McCabe C.,  2016

\bibitem[\protect\citeauthoryear{Wilms, Nowak, Pottschmidt, Pooley  \&
  Fritz}{Wilms et~al.}{2006}]{Wilms:2005xn}
Wilms J.,  Nowak M.~A.,  Pottschmidt K.,  Pooley G.~G.,   Fritz S.,  2006,
  \mn@doi [Astron. Astrophys.] {10.1051/0004-6361:20053938}, 447, 245

\bibitem[\protect\citeauthoryear{Yuan \& Zhang}{Yuan \&
  Zhang}{2014}]{Yuan:2014rca}
Yuan Q.,  Zhang B.,  2014, \mn@doi [JHEAp] {10.1016/j.jheap.2014.06.001}, 3-4,
  1

\bibitem[\protect\citeauthoryear{Zhang, Xin, Fu, Zhou, Yan, Liu  \&
  Zhang}{Zhang et~al.}{2016}]{Zhang:2016tcf}
Zhang P.~F.,  Xin Y.~L.,  Fu L.,  Zhou J.~N.,  Yan J.~Z.,  Liu Q.~Z.,   Zhang
  L.,  2016, \mn@doi [Mon. Not. Roy. Astron. Soc.] {10.1093/mnras/stw567}, 459,
  99

\bibitem[\protect\citeauthoryear{Zhou, Liang, Huang, Li, Fan, Feng  \&
  Chang}{Zhou et~al.}{2015}]{Zhou:2014lva}
Zhou B.,  Liang Y.-F.,  Huang X.,  Li X.,  Fan Y.-Z.,  Feng L.,   Chang J.,
  2015, \mn@doi [Phys. Rev.] {10.1103/PhysRevD.91.123010}, D91, 123010

\bibitem[\protect\citeauthoryear{Zhu, Li  \& Morris}{Zhu
  et~al.}{2018}]{Zhu:2018pkm}
Zhu Z.,  Li Z.,   Morris M.~R.,  2018

\bibitem[\protect\citeauthoryear{van Haaften, Nelemans, Voss, Wood  \&
  Kuijpers}{van Haaften et~al.}{2012a}]{vanHaaften:2011iy}
van Haaften L.~M.,  Nelemans G.,  Voss R.,  Wood M.~A.,   Kuijpers J.,  2012a,
  \mn@doi [Astron. Astrophys.] {10.1051/0004-6361/201117880}, 537, A104

\bibitem[\protect\citeauthoryear{van Haaften, Nelemans, Voss  \& Jonker}{van
  Haaften et~al.}{2012b}]{vanHaaften:2012zn}
van Haaften L.~M.,  Nelemans G.,  Voss R.,   Jonker P.~G.,  2012b, \mn@doi
  [Astron. Astrophys.] {10.1051/0004-6361/201218798}, 541, A22

\bibitem[\protect\citeauthoryear{van Haaften, Voss  \& Nelemans}{van Haaften
  et~al.}{2012c}]{vanHaaften:2012qc}
van Haaften L.~M.,  Voss R.,   Nelemans G.,  2012c, \mn@doi [Astron.
  Astrophys.] {10.1051/0004-6361/201118067}, 543, A121

\bibitem[\protect\citeauthoryear{van Haaften, Nelemans, Voss, Toonen,
  Portegies~Zwart, Yungelson  \& van~der Sluys}{van Haaften
  et~al.}{2013}]{vanHaaften:2013uuo}
van Haaften L.~M.,  Nelemans G.,  Voss R.,  Toonen S.,  Portegies~Zwart S.~F.,
  Yungelson L.~R.,   van~der Sluys M.~V.,  2013, \mn@doi [Astron. Astrophys.]
  {10.1051/0004-6361/201220552}, 552, A69

\bibitem[\protect\citeauthoryear{van Haaften, Nelemans, Voss, van~der Sluys  \&
  Toonen}{van Haaften et~al.}{2015}]{vanHaaften:2015wta}
van Haaften L.~M.,  Nelemans G.,  Voss R.,  van~der Sluys M.~V.,   Toonen S.,
  2015, \mn@doi [Astron. Astrophys.] {10.1051/0004-6361/201425303}, 579, A33

\makeatother
\end{thebibliography}



\appendix


\section{Galactic binaries and Jets}

\subsection{Eddington accretion rate}
\label{sec:luminosity}
The Eddington luminosity, $L_\mathrm{Edd}$, defined as the luminosity
at which the radiation pressure due to Thomson scattering on the electrons equals the gravitational force on the protons (which dominate the mass), is
\begin{align}
  \begin{split}
  	L_\mathrm{Edd} 	&= \frac{4\pi c G M m_p}{\sigma_T} \\
    				&\approx 1.26\times 10^{38}
                    \left(\frac{M}{M_\odot}\right)
                    \mathrm{\,erg\,s^{-1}},
  \end{split}
\end{align}
with $M$ the mass of the accreting object, $m_p$ the proton mass
and $\sigma_T \approx 0.665\mathrm{\,barn}$ the Thomson cross section.

The mass accretion rate can be related to the bolometric luminosity 
through the available gravitational potential energy (e.g.~\cite{1992apa..book.....F})

\begin{equation}
	L_\mathrm{Bol} = \epsilon \frac{G M_\mathrm{acc} \dot M}{R_\mathrm{acc}}.
\end{equation}
Here $R_\mathrm{acc}$ the radius of the accretor. In case of a 
a NS it is $R = 10\mathrm{\,km}$.
For a BH $R_\mathrm{acc} = 6 G M_\mathrm{acc} / c^2$,  3 times the Schwarzschild radius,
corresponding to the innermost stable orbit of a non-spinning ($a=0$) BH. 
For a maximally-rotating BH ($a=1$) it is 
$R_\mathrm{acc} = \frac 1 2 G M_\mathrm{acc} / c^2$\footnote{
For a $1.4\mathrm{\,M_\odot}$ NS this is $\sim 12\mathrm{\,km}$.} \citep{1989ApJ...341..948H, 1992apa..book.....F, Belczynski:2005mr}.
For efficient accretors, the parameter $\epsilon$ gives the conversion efficiency of gravitational potential energy into luminosity and is $1\,(0.5)$ for surface (disk) accretion, i.e.~ for NSs (BHs) \citep{Shakura:1976xk,
1992apa..book.....F, Belczynski:2005mr}.
For BHs the luminosity can thus be written as 
$L_\mathrm{bol}=\eta \dot M c^2$, where $\eta = \epsilon \frac{G M_\mathrm{acc}}{R_\mathrm{acc}} \sim \epsilon / 6 \sim 0.1$ in 
Eq.~(\ref{eq:Lbol}) (for non-rotating BHs).
For NSs we also assume $\eta\sim 0.1$
in this work,
since the propeller mechanism effectively works 
as an event horizon.

The Eddington accretion rate can now be written as
\begin{align}
  \begin{split}
  	\dot M_\mathrm{Edd}	&= \frac{L_\mathrm{Edd}}{\eta c^2} \\
    					&\approx 2.22\times 10^{-8}
                        \left(\frac{\eta}{0.1}\right)^{-1}
                        \left(\frac{M}{M_\odot}\right)
                        \mathrm{\,M_\odot\,yr^{-1}}.
  \end{split}
\end{align}

\subsection{Individual sources}
\label{sec:individual_sources}
If Eq.~(\ref{eq:Npos}) would hold for all accreting binaries
including those that are not in the low/hard state-- 
i.e.~if all jets would be composed of a cold electron-positron
pair plasma-- positrons would be vastly overproduced in the Galaxy. 
However, different jets are expected to have different properties.
For instance, transient jets are expected to be more relativistic than
steady jets \citep{Fender:2004gg}. In this paper we assume
that low-energy positrons are predominantly produced
in steady-jets. Below we discuss a few interesting sources
that are in the low/hard or quiescent regime
in more detail. We recall that the $2\sigma$ narrow-line
sensitivity is $5.7\times10^{-5} \sqrt{10^{6}/T_\mathrm{obs}}
\mathrm{\,ph\,cm^{-2}\,s^{-1}}$ \citep{Siegert:2016ijv}.

\subsection*{Neutron stars}
Here we discuss the two most promising transitional millisecond pulsars,
J1023-0038 and XSS J12270-4859, in more detail. 
Two more (candidate) tMSP exists. The tMSP J1824-2452I \citep{Linares:2013ewa, Papitto:2013hza} is in a globular cluster and is more distant
than the aforementioned two systems. Also, it is more likely in a high/soft state. For this reason we omit it from further discussion.
In addition, J1544.6-1125 \citep{Bogdanov:2015xda} is a candidate
tMSP with an unconstrained distance. Interestingly, it is associated
with a \Fermi~$\gamma$--ray source.

\subsubsection{PSR J1023+0038}
This transitional millisecond pulsar has
a low-mass accretion rate and is nearby ($1.37 \mathrm{\,kpc}$) 
\citep{Archibald:2009zb,Archibald:2014nda,Deller:2014yca}. For its estimated mass-accretion rate
of $10^{-1}$--$10^{-1}\mathrm{\,\dot{M}_\odot \,yr^{-1}}$
we obtain a jet power of $\sim 6\times 10^{32}$--
$6\times 10^{34}\mathrm{\,erg\,s^{-1}}$.
This would correspond to a 511 keV flux of 
$\sim 1\times 10^{-7}$--
$1\times 10^{-5}\mathrm{\,ph\,s^{-1}\,cm^{-2}}$.
In the most optimistic scenario, a narrow line
from this source could be observable by INTEGRAL/SPI 
with sufficient exposure, however 
the exposure at the source position, $(\ell, b) = (243.5^\circ, 45.8^\circ)$, is only $\sim 25\mathrm{\,ks}$
(see Sect.~\ref{sec:MWGC}).

J1023+0038 has also been identified as a $\gamma$-ray source in the second \Fermi-LAT source catalog, where it is classified
as an associated pulsar
\citep{Fermi-LAT:2011yjw}. In the FL8Y 
it is listed as a $\gamma$-ray detected pulsar.

\subsubsection{XSS J12270-4859}
This LMXB was identified as a tMSP \citep{Bassa:2014hoa, Roy:2014cwa} and also is listed as a pulsar candidate in the 3FGL \citep{Acero:2015hja} and 
as \Fermi-LAT $\gamma$-ray detected pulsar in the FL8Y. 
It is expected to be at a similar distance as J1023+0038 and has a slightly higher X-ray luminosity \citep{Papitto:2014rua,Deller:2014yca}. As such, we expect a slightly higher 511 keV flux from this source than from J1023+0038. 
Furthermore, there is a
high exposure of $\sim 3\times 10^7\mathrm{\,s}$ at the source location, therefore this source is potentially already detectable with INTEGRAL/SPI.
This makes XSS J12270-4859 arguably the most promising candidate.

\subsection*{Black holes}
\subsubsection{Sagittarius A* }
Sagittarius A* (Sgr A*) is the supermassive BH at
the center of our Galaxy. It has a mass of $\sim 4\times 10^6\mathrm{\,M_\odot}$. 
Its bolometric luminosity of $\sim 10^{36}\mathrm{\,erg\,s^{-2}}$
is far below the Eddington luminosity \citep{Prantzos:2010wi}.
Taking this luminosity and using Eqs.~(\ref{eq:Pj_BH}, \ref{eq:Npos}, \ref{eq:N511}) as we did for Galactic binaries would lead to an over production of positrons by about three orders-of-magnitude. 
When one also takes into account the mass dependence
of the radio--X-ray correlation through the fundamental plane 
the jet power would increase by an additional three orders of magnitude \citep{Merloni:2003aq, Falcke:2003ia}.
If binary jets are responsible for the 511 keV emission through
the mechanism explained in the main text, their jets should differ
from that of Sgr A*, unless NSs are primarily responsible for the
positrons and their jet composition differs from that of BH jets.

\subsubsection{Cygnus X-1}
Given the estimate for the jet power from \cite{Gallo:2005tf}, 
the high-mass X-ray binary Cyg X-1 should have been observed as a 511 keV point source if 
the jet where composed of a cold, pair-dominated plasma and if the outflowing
positrons would annihilate in the vicinity of the source, as proposed in this work. 
However, upper limits on the 511 keV line flux from 
\citep{Knodlseder:2005yq, Jourdain:2011we, Siegert:2015knp} are in mild tension with this scenario.

The tension can be relieved when the state of Cyg X-1 is taken into account, 
since it spends a significant fraction in the high/soft state \citep{Wilms:2005xn}. 
Moreover, if the leptons are more energetic than in our estimate the observed jet power requires fewer particles.
In case of Cyg X-1 it is likely that the jet kinetic energy is carried by 
protons \citep{Gallo:2005tf, Heinz:2005jc}. For a one-to-one ratio of protons-to-electrons
the number of positrons injected, and correspondingly the number of 511 keV
photons is reduced by a factor $m_e/m_p$ in which case the upper limits are
again satisfied.

\subsubsection{A0620-00}
A0620-00 is a quiescent BH ($M=11\mathrm{\,M_\odot}$) radiating 
at a luminosity corresponding
to $10^{-9}\text{--}10^{-8}\,L_\mathrm{Edd}$ and located at a distance
of $1.2\mathrm{\,kpc}$ \citep{Gallo:2006dz}.

Using the assumptions and expressions in Sect.~\ref{sec:bin_pos} we estimate 
the 511 keV flux from this source to be 
$\sim 10^{-5}\mathrm{\,ph\,cm^{-2}\,s^{-1}}$.
With a current exposure of $\sim 3\mathrm{\,Ms}$ at the source such a flux is currently undetectable, albeit not by large margin. More observation time with INTEGRAL/SPI could potentially reveal a 511 keV signal from this source.
Moreover, a future mission such as e-ASTROGAM \citep{DeAngelis:2016slk} might be able
to detect this signal. The expected sensitivity for a $3\sigma$ detection
of a narrow 511 keV line by e-ASTROGAM is 
$4.1\times 10^{-6}\mathrm{\,ph\,cm^{-2}\,s^{-1}}$ for $10^6\mathrm{\,s}$ of 
effective observation time. Taking the sensitivity to scale with the square
root of time and using a field-of-view of $2.5\mathrm{\,sr}$, a flux of 
order $5\times 10^{-6}\mathrm{\,ph\,cm^{-2}\,s^{-1}}$ should be attainable within about a month of total mission time. 

If A0620-00 is detected as a source of 511 keV emission this would provide
supporting evidence for the hypothesis that the 511 keV bulge emission comes
from dim BH X--ray binaries.

\subsubsection{47 Tuc X9}
The ultracompact XRB 47 Tuc X9 resides in the globular cluster 47 Tucanae
and it has recently been argued that it potentially has a BH accretor, if so
being the first of its kind \citep{Bahramian:2017mbs}.
If the accretor is not a BH, this system is most likely
a tMSP, as suggested by its position in the radio--X-ray correlation plane.
The X-ray luminosity in the 1--10 keV band is a 
$\mathrm{few}\times 10^{33}\mathrm{\,erg\,s^{-1}}$ \citep{Bahramian:2017mbs}
corresponding to 
a total hard X-ray luminosity of $\sim 10^{34}\mathrm{\,erg\,s^{-1}}$
($\sim 10^{-5}\,L_\mathrm{Edd}$).
Performing an estimate for the positron annihilation signal following
Sect.~\ref{sec:bin_pos} we estimate the jet power to be 
$\sim 10^{35}\mathrm{\,erg\,s^{-1}}$ and the 511 keV line flux $\sim 10^{-5}\mathrm{ph\,cm^{-2}\,s}$, using a distance
of $4.53\mathrm{\,kpc}$.
In the most optimistic scenario, 47 Tuc X9 is already a candidate 511 keV source for INTEGRAL/SPI.
If the BH nature is confirmed, a detection would be direct evidence that BH-UCXBs can inject low-energy positrons into the bulge. However, any 511 keV signal from 47 Tuc X9 will be degenerate with that of the GC as a whole.

\section{Detectability of dim 511-keV point sources with IBIS.}
\label{sec:IBIS}
In this section we estimate the potential of INTEGRAL/IBIS 
to study a dim point source origin of the 511 keV emission.
It is shown that the presence of a 511-keV-point-source population 
similar to the one discussed in the main text can in theory
be revealed through analyses of the photon count statistics.
We stress however that simplifying assumptions have been 
made in the treatment of the coded-mask instrument IBIS and that a 
more detailed analysis of instrumental effects would be necessary in order
to assess more realistically the sensitivity of such a search. Such a
simulation is beyond the scope of this work.

\subsection{Mock sky images}
We analyze mock sky images in a region-of-interest (ROI) of $40^\circ\times40^\circ$ centered
on the inner-Galaxy to study the impact of a dim point-source
population in the bulge on the photon count statistics.

\subsubsection*{Instrumental modeling}
Sky-images are constructed in the energy $491\text{--}531\mathrm{\,keV}$, similar to the IBIS analysis of the inner-Galaxy as performed by \cite{DeCesare:2011gc}. For the field-of-view (FoV) we consider the fully-coded field-of-view 
of $9^\circ\times9^\circ$ 
(note that the larger partially coded field of view is $29^\circ\times29^\circ$) 
\citep{Goldoni:1997ti,Goldwurm:2003vx}).
The effective area of IBIS is taken to be 
$A_\mathrm{eff}=54.2\mathrm{\,cm^2}$ and the exposure towards
the inner-Galaxy $T_\mathrm{obs} = 10^7\mathrm{\,s}$.
Finally, we include a term $\epsilon_\mathrm{eff}=0.75$ 
to capture the imaging efficiency \citep{DeCesare:2011gc}.
Given some photon flux, $\phi$, the total number of source
photons is:

\begin{equation}
	N_\mathrm{S} = 
    \phi A_\mathrm{eff}\epsilon_\mathrm{eff}T_\mathrm{obs}.
\end{equation}

\subsubsection*{Source and background fluxes}
We model the source and background fluxes
following the model from \cite{Siegert:2015knp} for
the all-sky distribution of 511 keV emission.
The spatial model of \cite{Siegert:2015knp} contains 
3 model components in our ROI (see their Table 2): a bulge, a disk and a central source. The bulge consists out of a broad and narrow
component which we assume contain respectively 72\% and 28\% of the total bulge flux \citep{PoS(INTEGRAL2014)054}.

In the energy window $491\text{--}531\mathrm{\,keV}$ 
the spectrum consists out of three components:
(i) the 511 keV line due to p-Ps annihilation; (ii) a continuum,
starting from 511 keV\footnote{Note that non of the three photons
can have an energy above 511 keV due to momentum conservation.},
due to o-Ps annihilation which has a 3 photon final-state; and
(iii) a background $\gamma$-ray continuum which is modeled as 
a powerlaw. The background $\gamma$-ray continuum for each spatial
component is modeled as:
\begin{align}
  \left.\frac{d\Phi}{dE}\right|_{i} = A_i 
    \left(\frac{E}{511 \mathrm{\,keV}}\right)^{-1.7}
    \mathrm{\,ph\,cm^{-2}\,s^{-1}\,keV^{-1}},
\end{align}
where the normalizations $A_i$ are taken from Table 5 of
\cite{Siegert:2015knp}. 
In Table 5 \cite{Siegert:2015knp} also provide the flux of photons
from o-Ps and p-Ps from the various spatial components. It turns out 
that in our ROI $\sim 90\%$ of the astrophysical $\gamma$-rays are due
to positronium annihilation.

Finally, we assume a total positron injection rate of 
$\dot N_\mathrm{e^+} = 2\times10^{43} \mathrm{\,s^{-1}}$ in the bulge. 
Since 511 keV annihilation-line spectroscopy points towards
a positronium fraction of $\sim 100\%$, we assume that 25\% of these photons
annihilate to the 511 keV emission and 75\% through 
the 3-photon final state. The total photon flux from positronium in the bulge in our energy window then becomes:
\begin{equation}
  \dot N_\mathrm{\gamma,\,Ps} = \frac{11}{4}\dot N_\mathrm{e^+}.
\end{equation}

In terms of the spatial distribution, we assume that all point sources
trace the bulge and Galactic-Center component from the \cite{Siegert:2015knp} model, and that each source is located $8.3\mathrm{\,kpc}$ away. 

Finally, sources are randomly sampled from the luminosity
functions in Figs \ref{fig:Lx_NS} and \ref{fig:Lx_BH} assuming
a BH:NS ratio of 1:5.
The positron yield is then
calculated following Eq.~\ref{fig:positrons} until 
the total injection rate saturates $2\times 10^{43}\mathrm{\,e^+\,s^{-1}}$. Spatial positions 
are drawn from the templates of \cite{Siegert:2015knp}.
Source fluxes are smoothed by a gaussian kernel with $\sigma=0.25^\circ$ to represent the IBIS PSF and we add
Poisson noise to the final image.

\subsubsection*{Instrumental modeling}
The IBIS instrument has a large internal background due to 
cosmic-rays interacting in the detector \citep{Lebrun:2003aa}.
From Fig.~10 of \cite{Lebrun:2003aa} we can estimate the background
in the 491--531 keV range to be $\sim 7 \mathrm{\,ph\,s^{-1}}$. 
We will henceforth assume a background of $10 \mathrm{\,ph\,s^{-1}}$
similar to \cite{DeCesare:2011gc}.

We assume that the instrumental background is spread out uniformly 
over the $9^\circ\times9^\circ$ FoV and follows poison statistics.
The total instrumental background thus becomes 
$\sim 405\mathrm{\,ph\,s^{-1}\,sr^{-1}}$.

\subsubsection*{Sky images}
\begin{figure*}
	\includegraphics[width=\textwidth]{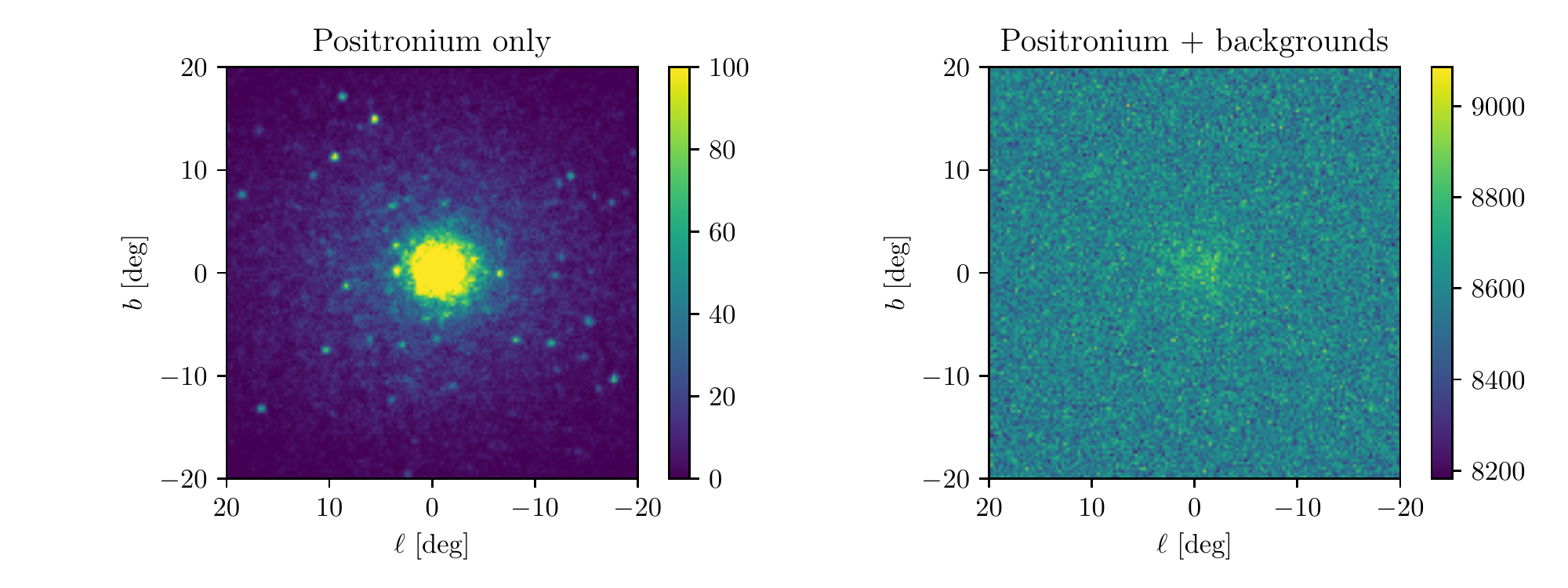}
    \caption{
    \textit{Left panel:} example of the flux from
    a bulge point-source population producing the 511 keV emission
    in from the inner-Galaxy. Photon counts are for an exposure
    10 Ms and in pixels of $0.05^\circ\times 0.05^\circ$.
    \textit{Right panel:} the same as the left panel,
    but now including all backgrounds.
    }
    \label{fig:511map}
\end{figure*}

\begin{figure}
	\includegraphics[width=\columnwidth]{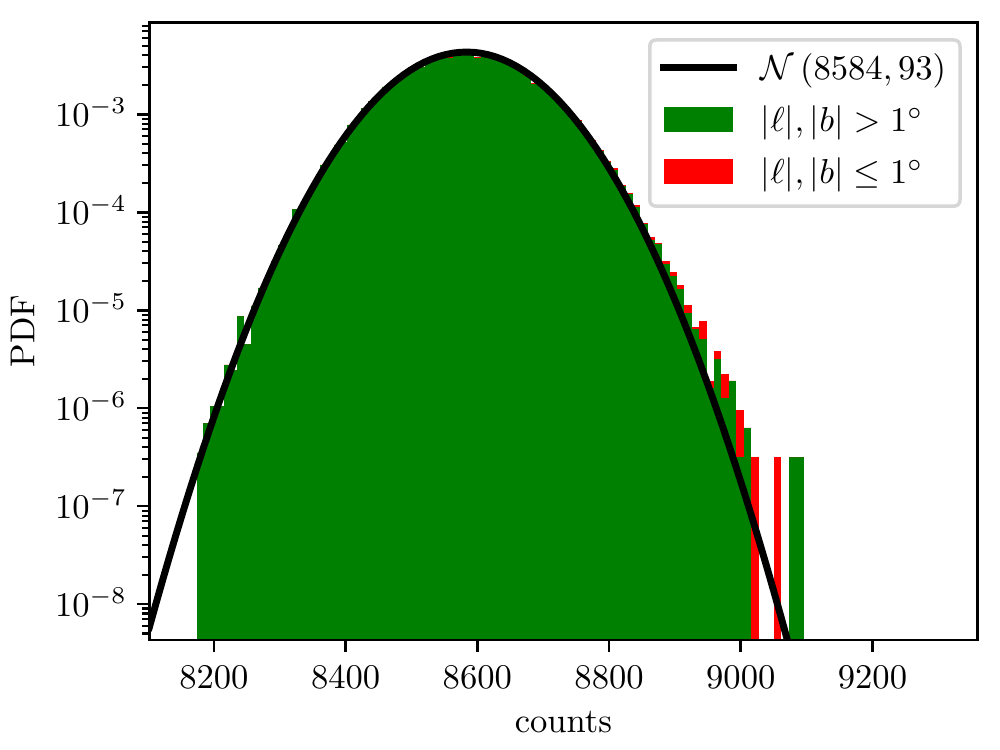}
    \caption{Photon distribution corresponding to the
    right panel of Fig.~\ref{fig:511map}.
    The instrumental background was assumed to be poissonian.
    Evidently, there is only a minor non-poissonian tail 
    in the photon count contribution due to
    the unresolved point sources.
    Pixel counts from the Galactic Center (red, $|\ell|, |b|<1^\circ$)
    and the remainder of the ROI (green) are displayed separately.
    }
    \label{fig:ph_distribution}
\end{figure}

We combine the instrumental background with the astrophysical fluxes
to produce sky images. In Fig.~\ref{fig:511map} we show
the signal due to positronium (left panel) and the total 
flux (right panel) in our ROI for an exposure of 10 Ms. As can
be seen, the instrumental background completely dominates the image. Figure \ref{fig:ph_distribution} shows the pixel count
distribution for the right panel of Fig.~\ref{fig:511map}.
Clearly, there is an almost negligible deviation from a Poisson 
distribution due to unresolved points sources.
\subsection{Wavelet analysis}
In order to see if we can identify the presence of point sources
in the mock data we apply a wavelet transform to the mock data.
This method is similar to what was
applied by \cite{Bartels:2015aea} to search for unresolved point sources in the \Fermi-LAT data of the inner-Galaxy.
The wavelet analysis is a powerful tool to identify sudden spatial fluctuations in intensity, in this case due to the presence of point sources, on top of a slowly varying background.

We apply Mexican-hat-wavelet-family member one \citep[e.g.~][]{GonzalezNuevo:2006pa} with a scale of $0.4^\circ$.
We quantify a signal-to-noise ratio $\mathcal{S}$ similar to what is
done by \cite{Bartels:2015aea}. For more details of the analysis
we refer the reader to this paper. 
\begin{figure*}
	\includegraphics[width=\textwidth]{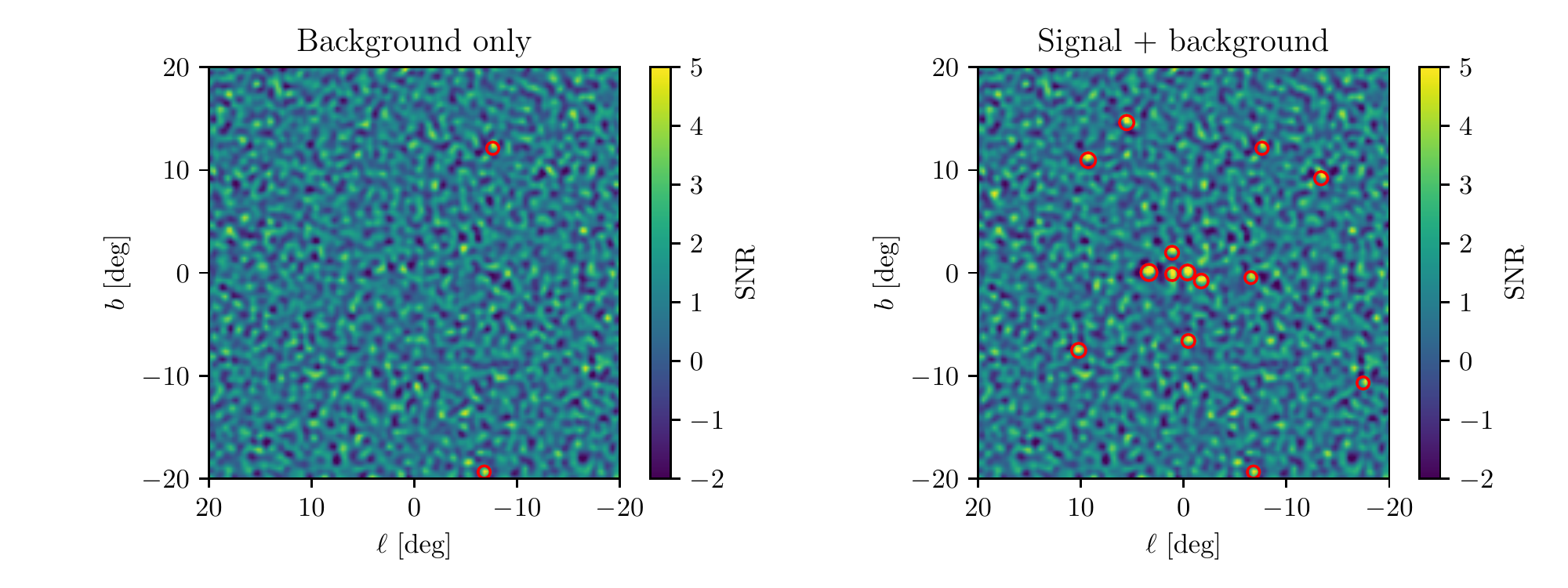}
    \caption{Wavelet transformed map of a background only
    image (left panel) and an image containing a dim point source
    population (right image). Red circles indicate sources with
    a signal-to-noise ratio above 5.
    }
    \label{fig:wavemap}
\end{figure*}
In Fig.~\ref{fig:wavemap} we show the wavelet transformed skymaps
for background only (left) and background plus sources (right).
All peaks with  $\mathcal{S}\geq 5$ have red circles, with larger
circles corresponding to larger signal-to-noise ratios.
\begin{figure}
	\includegraphics[width=\columnwidth]{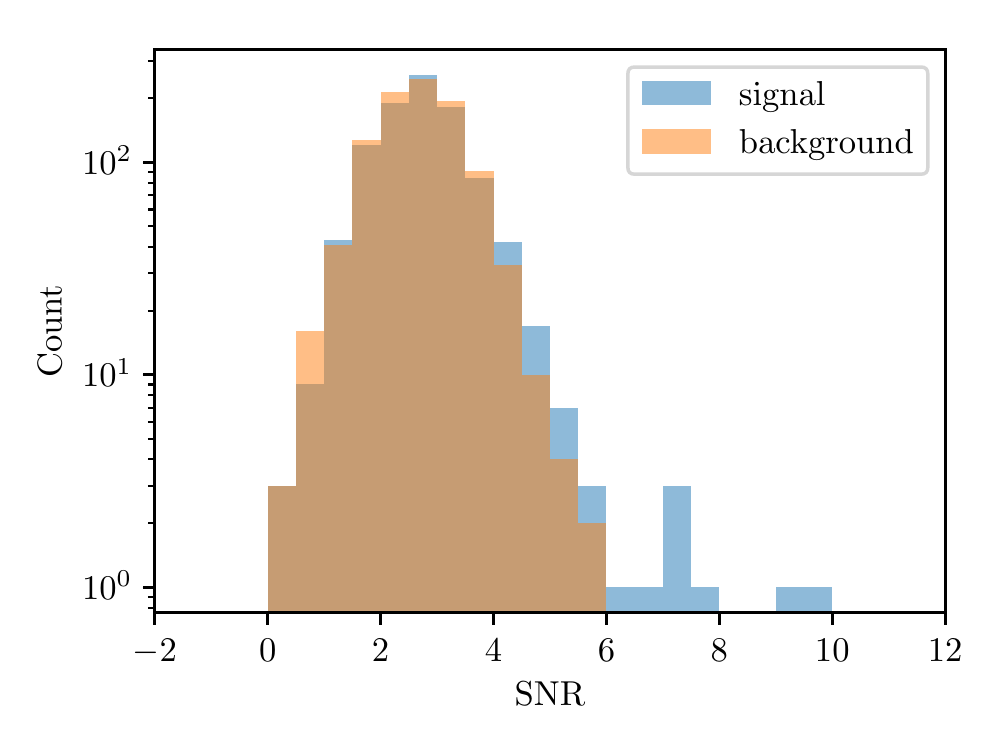}
    \caption{Number of wavelet peaks of a given signal-to-noise
    ratio in case of background only (orange) and background
    plus sources (blue).
    }
    \label{fig:w_snr}
\end{figure}
In Fig.~\ref{fig:w_snr} we
show the total number of wavelet peaks with a given $\mathcal{S}$
for the background only map (orange) and for background plus point
sources (blue).
A clear enhancement in the number of wavelet peaks with $\mathcal{S}\gtrsim 5$ can be seen in the presence of
point sources.

This analysis has shown that in principle one can identify
a dim point source population, similar to the one studied in the
main text of this paper, with dedicated analyses of photon
statistics using wavelets, despite the large instrumental 
backgrounds in IBIS. We note that the wavelet technique is insensitive
to fluctuations on scales much larger than the wavelet scale,
and can therefore still work if backgrounds are non-uniform.
Finally, the wavelet technique can corroborate the presence of power on small scales, i.e.~
the presence of point sources. Nevertheless, 
it will be challenging to 
reconstruct more details of any such point source population.

\bsp	
\label{lastpage}
\end{document}